\documentclass[final,1p,times]{elsarticle}

\usepackage[utf8]{inputenc}
\usepackage{lineno}
\usepackage{amsmath}
\usepackage{caption}
\usepackage{booktabs}
\usepackage{amsfonts}
\usepackage{algorithm}
\usepackage[noend]{algpseudocode}
\usepackage{longtable}
\usepackage{multicol}
\usepackage{rotating}
\usepackage{placeins}
 \usepackage{multirow}
\usepackage{hyperref}

\usepackage{color}
 
\newcommand\blue[1]{{\color{black}#1}}
\newcommand\green[1]{{\color{black}#1}} 
\newcommand\magenta[1]{{\color{black}#1}} 
\definecolor{OliveGreen}{rgb}{0,0.6,0}

\DeclareUnicodeCharacter{220A}{\ensuremath{\ \varepsilon}}

\journal{Waste Management}

\begin{document}

\begin{frontmatter}



\title{Exact and heuristic approaches for multi-objective garbage accumulation points location in real scenarios}


\author[address1]{Diego Gabriel Rossit\corref{mycorrespondingauthor}}
\cortext[mycorrespondingauthor]{Corresponding author}
\ead{diego.rossit@uns.edu.ar}
\author[address2]{Jamal Toutouh}
\author[address3]{Sergio Nesmachnow}

\address[address1]{INMABB, Department of Engineering, Universidad Nacional del Sur (UNS)-CONICET, Alem Av. 1253, B8000CPB Bah\'{i}a Blanca, Argentina}
\address[address2]{MIT Computer Science and Artificial Intelligence Laboratory, Massachusetts Institute of Technology. 32 Vassar St, Cambridge, MA 02139, USA}
\address[address3]{Faculty of Engineering, Universidad de la Rep\'{u}blica, 565 Julio Herrera y Reissig Av., 11300 Montevideo, Uruguay}

\begin{abstract}
Municipal solid waste management is a major challenge for nowadays urban societies, because it accounts for a large proportion of public budget and, when mishandled, it can lead to environmental and social problems. This work focuses on the problem of locating waste bins in an urban area, which is considered to have a strong influence in the overall efficiency of the reverse logistic chain. This article contributes with an exact multiobjective approach to solve the waste bin location in which the optimization criteria that are considered are: the accessibility to the system (as quality of service measure), the investment cost, and the required frequency of waste removal from the bins (as a proxy of the posterior routing costs). In this approach, different methods to obtain the objectives ideal and nadir values over the Pareto front are proposed and compared. Then, a family of heuristic methods based on the PageRank algorithm is proposed which aims to optimize the accessibility to the system, the amount of collected waste and the installation cost. The experimental evaluation was performed on real-world scenarios of the cities of Montevideo, Uruguay, and Bah\'{i}a Blanca, Argentina. The obtained results show the competitiveness of the proposed approaches for constructing a set of candidate solutions that considers the different trade-offs between the optimization criteria.
\end{abstract}


\end{frontmatter}

\noindent
\textbf{Copyright notice:} This article has been accepted for publication in the Waste Management journal. Cite as: Rossit DG, Toutouh J, Nesmachnow S. Exact and heuristic approaches for multi-objective garbage accumulation points location in real scenarios. \textit{Waste Management}. 2020 Mar 15;105:467-481. doi: 10.1016/j.wasman.2020.02.016. Epub 2020 Mar 2. PMID: 32135468.
\section{Introduction}
\label{sec:intro}

In the last years, the concept of \textit{smart cities} has emerged to represent the application of decision support systems to improve the efficiency of public services in urban agglomerations. \citet{harrisonX2010foundations} \green{recognize} a city to be ``smart'' if it is instrumented, i.e., \green{uses} real-time real-world data, interconnected, i.e., integrates the gathered data and communicates it among various city services, and intelligent, i.e., \green{includes} modeling, optimization, and visualization in the operational decision making process. Municipal Solid Waste (MSW) management is a promising area for applying intelligent smart cities characteristics to improve the decision making process~\citep{deX2017technologies}, not only because it represents one of the major expenses in the budget of local governments but also because, when mishandled, is associated with severe social and environmental impacts~\citep{hoornweg2012waste}. 

Among the several decisions that a MSW system involves, it is the design of the waste collection network, which is the entry point to the system. 
This problem involves finding the best locations for community waste bins, that are arranged in special places known as \textit{Garbage Accumulation Points} (GAPs), in an urban area while optimizing some relevant criteria that is usually related to the cost of the system, to the Quality of Service (QoS) provided to the citizens or to both of them in a multiobjective fashion. The number of GAPs, their distribution in the field, the type and capacity of the bins used in each point, and the frequency of waste removal from these bins are major conditioning factors of the overall efficiency of the MSW system. Not only GAPs location and configuration strongly influence the operational routing cost of waste collection from bins to the disposal sites\green{,} but also have an impact on more strategic levels of this reverse logistic chain, e.g., the designed capacity of intermediate and processing facilities~\citep{chaudharyX2019gis}. 

Besides its relevance, locating GAPs is not a trivial problem. Firstly, because in terms of computational complexity, the GAP location problem is an extension of the Capacitated Facility Location Problem (CFLP), which was proven to be NP-hard~\citep{cornuejolsX1991comparison} 
\magenta{(i.e., at least as hard as problems for which no efficient algorithms, that executes in polynomial time with respect to the size of the problem, have been devised. This class characterizes those problems that are difficult to solve computationally)}.
Secondly, because \green{of} the conflicting goals which are involved \green{since waste bins are considered \textit{semi}-obnoxious facilities~\citep{tralhaoX2010multiobjective}. On the one hand, citizens that live near to the bins can suffer different environmental costs, such as noise pollution, bad smell, visual pollution, and traffic congestion from collection vehicles~\citep{flahautX2002locating}. Moreover, this can also affect the selling price of the nearby buildings~\citep{di2014integrationpilotcase}. However, on the other hand, citizens that live too far from the bins have to carry their waste for long distances, what can affect the accessibility to the system. When citizens are unwilling to incur in this transport cost, waste might be dumped in unsuitable places~\citep{parrotX2009municipal} which have to be removed by the authorities (incurring in additional expenses)}. This conflicting relation is associated with the ``Not in my Back Yard'' 
response to undesirable facilities: the citizens who accept that these facilities are placed nearby are fewer than the ones who admit that they should be placed somewhere~\citep{lindell1983close}. However, not only environmental costs take part in the trade-off. \citet{ghianiX2012capacitated} stated that despite their eagerness to have a GAP as close as possible, citizens are also willing to have the smaller tax burden that guarantees the service.

\blue{
Regarding the research objective and motivation,
this article contributes to the state-of-the-art related literature with a new multiobjective \green{model} that simultaneously considers waste bins installment cost and QoS provided to the citizens together with the reduction of the number of required visits to empty the bins along the time horizon (i.e., collection frequency), which is an important contributor to the posterior routing cost of the waste collection~\citep{greco2015drivers}. The motivation of the inclusion of this last criteria is that, although is not a common feature of the state-of-the-art GAPs location literature, posterior routing costs are highly relevant in MSW system overall expenses~\citep{das2015optimization}. This situation is even more important in some of the case studies presented in this work, where decision-makers experience especially large transport costs~\citep{broz2018argentinian}. This formulation is solved with an exact approach based on the augmented $\varepsilon$-constraint method~\citep{mavrotas2013improved}. For this purpose, the first stage of this method has to be modified for dealing with this computationally challenging problem. Finally, a heuristic approach based on the web ranking PageRank algorithm~\citep{Langville2011} is proposed to obtain fast solutions for a GAP location problem}. \magenta{The main focus of the research is studying computational methods for the GAP location problem, in order to provide efficient and flexible support tools to help authorities in the decision-making process, especially in emerging countries. Efficient computational methods allows studying multiple alternatives for GAP location, making possible the analysis of dynamic situations that may arise in day-to-day operation of MSW systems (for example, blocked roads, non-availability of specific locations for GAP installation, complains of citizens for the current location of bins, variations on the collection fleet). Using efficient computational methods, these situations can be studied in advance, or even almost in real-time, in order to provide accurate modifications to be implemented in the city}.

The article is organized as follows. Section~\ref{sec:review} describes the work that have already been performed in this field. Section~\ref{sec:exact_multiobjective} presents a multiobjective exact approach to solve the GAP location problem. Section~\ref{sec:heuristic} presents a family of PageRank heuristics to solve a similar GAP location problem. Section~\ref{sec:experimental_evaluation} reports the experimental evaluation of the proposed approaches on real-world instances. Finally, Section~\ref{sec:conclusion} presents the conclusions of the research and formulates the main lines for future work. 

\section{Literature review}
\label{sec:review}

Several authors have addressed this tricky problem. \citet{bautista2006modeling} represented the problem using two different approaches: a minimal set covering problem and a maximum satisfiability problem. The goal is to locate the minimal number of collection points so that each dwelling has a collection point within a certain threshold distance by means of metaheuristic algorithms in scenarios of Barcelona, Spain. The same problem was solved by~\citet{ghianiX2012capacitated}, who proposed a constructive heuristic and validated the results by solving small instances with CPLEX. Later in \citet{ghianiX2014impact}, this heuristic was modified to bound posterior routing costs, i.e., bins that require different kind of vehicles to be collected are not allowed to be located in the same GAP. Both works solved scenarios of Nard\`{o}, Italy. \citet{di2014integrationpilotcase} proposed a two-phase heuristic. First, the GAPs location problem is solved and in a second phase the sizes of the bins to be assigned to the GAPs is determined. A distinctive characteristic of this model is that GAPs can be located at any point of the street urban network and not at predefined positions as is common in similar problems. Studies exploring the sensitivity of the necessary number of GAPs when the allowable threshold distance between a dwelling and its assigned GAP varies where performed in L'Aquila, Italy. \citet{rathoreX2019location} proposed a single objective mixed integer model to locate GAPs while minimizing the installment cost in Bilaspur, India.

Several works consider multiple criteria \green{for} optimizing GAP location. \citet{tralhaoX2010multiobjective} solved a multiobjective variant of the problem in an area of Coimbra (Portugal) considering four different objectives to minimize: the cost of the bin network, average distance between dwellings and assigned GAPs, and number of citizens within the ``push'' and ``pull'' thresholds of an open GAP. These last two objectives are related with the aforementioned semi-obnoxiousness of bins. \green{The authors} proposed two approaches for finding efficient solutions: weighted sum and goal programming (minimizing the distance to the ``ideal solution'', i.e., the infeasible solution that has the individual optimal value of each objective). \citet{coutinhoX2012bi} solved a similar problem, but only considering two objectives: the total investment cost and a novel ``dissatisfaction function'' that again considers the semi-obnoxiousness of bins. They applied the $\varepsilon$-constraint method to build the Pareto front of the problem. In \citet{rossitX2017application}, the augmented $\varepsilon$-constraint method~\citep{mavrotas2013improved} and weighted sum were used to solve a model that optimize the investment cost and the accessibility to the system in a simulated instance of Bah\'{i}a Blanca, Argentina. $\varepsilon$-constraint outperformed the weighted method in computing time, number of efficient solutions, and capacity of controlling the distribution of efficient points along the Pareto front of the problem. 

Other researchers have used Geographical Information System (GIS) technology for settling the location of bins in an urban area. \citet{aremu2012case} applied a GIS-based approach to solve a $p$-median problem to locate waste bins in Ilorin, Nigeria and analyzed the sensitivity of the solution regarding service coverage, public satisfaction, waste bin utility, costs associated with service provision, and emissions from collection vehicles. \citet{boskovic2015fast} located bins for a sector of Kragujevac, Serbia, using ArcGIS Network Analyst. \citet{erfaniX2018using} used the same program to solve case studies in Dhanbad, India, and Mashhad, Iran, respectively. First, the authors decided how to rearrange the current distribution of GAPs and then they decided how many bins to install. A similar approach was applied in Athens~\citep{karadimasX2005gis}. \citet{gallardoX2015methodology} made an important contribution by proposing an organized methodology to design a collection network divided in several stages: select the number of fractions in which waste is classified, select the system to be used to collect garbage, (e.g., a door-to-door system or community bins), select the GAP locations,  and select the types of bins that will be used. The authors applied their methodology in Castell\'{o}n, Spain using ArcGIS.

Although usually the problem of locating bins is solved separately from the design of collection routes, cost savings can be achieved with an integrated model~\citep{ghianiX2014operations}. \citet{chang2000siting} proposed a fuzzy multiobjective genetic algorithm to solve the recycling drop-off sites allocation and routing collection in Kaohsiung, Taiwan. They considered the percentage population served, the average walking distance for the residents to reach recycling drop-off stations, and the approximated routing distance of collection vehicles as fuzzy planning goals. \citet{hemmelmayrX2013models} proposed an integrated approach where the bins allocation problem is solved jointly with the routing schedule. The general framework is that the GAPs location problem is embedded in a Variable Neighbourhood Search (VNS) algorithm that determines the routing collection circuits. The authors used four overall algorithmic approaches with the aim of minimizing the total cost, i.e., the investment cost of the bins localization and the routing cost. The first two are hierarchical approaches that optimize in sequence the two parts of the problem, i.e., bins allocation and then vehicle routing, and vice versa. The third one is a integrated method that solves the overall problem by dynamically varying service frequencies and bin allocations. In the three approaches, the bin allocation part is solved with CPLEX. Finally, in the fourth approach the authors used a heuristic estimation for the bin allocation part. The general heuristic approach was validated by comparing the results of small instances solved with a mixed-integer programming model in CPLEX. Some articles propose solving the facility location and the routing sequentially. In \citet{linX2011model}, the GAPs location is first optimized with an integer programming model and then the routes are determined with an ant colony metaheuristic. The authors considered that the collection points have a temporal availability, i.e., generators only can carried their garbage during the collection site time window. The experimental evaluation was performed on an scenario of Taichung, Taiwan. \citet{ghianiX2014impact} also solved both problems sequentially for a case study in Nard\`{o}, Italy. \citet{erfaniX2017novel} conducted the same procedure in an scenario of the city of Mashhad, Iran.

\section{An exact multiobjective approach \green{for} GAP location}
\label{sec:exact_multiobjective}

\magenta{This section presents an exact multiobjective approach for solving the GAP location problem. The outline of the mathematical formulation is described in Section~\ref{subsec:formulation} and the resolution approach is described in Section~\ref{subsec:solution_approach}}.

\subsection{A comprehensive mathematical formulation}
\label{subsec:formulation}

The \green{proposed} model aims at proposing a bin network that:

\begin{enumerate}
\item minimizes the average required collection frequency of the containers. This criterion aims at assuring that the period of time between two consecutive visits of the collection vehicle to a bin for emptying their contents is as large as possible; 
\item minimizes the installation cost of bins;
\item minimizes the average distance between the generators and the assigned bins. This aims to enhance the accessibility of citizens to the collection network and, thus, is a metric of the QoS provided to the users that has been used in the related literature~\citep{coutinhoX2012bi,ghianiX2012capacitated,tralhaoX2010multiobjective}.
\end{enumerate}

The aforementioned criteria imply conflicting goals. If the average collection frequency is low, i.e., the period of time between every collection is relatively large, then the necessary storage capacity in GAPs (proportional to investment cost) will be large. Similarly, an overmuch comfortable assignment where generators can find GAP very near their home will imply large investment costs. This situation also will have a negative impact in the design of the collection routes of the containers due to the numerous bins that will be installed and are required to be visited~\citep{kaoX2002shortest}.

The mathematical formulation is presented as an Integer Programming (IP) model. The sets and parameters of the model are:
\begin{itemize}
\item A set $J$ of bin types. Each bin type has a given purchase price $cin_j$, capacity $cap_j$, and required space for its installation $e_j$.
\item A set $I$ of predefined potential locations where a GAP can be opened. Each potential location $i$ has an available space $Es_i$ for bins.
\item A set $P$ of (groups of) generators. The way these groups are conformed is explained \green{in} Section~\ref{sec:experimental_evaluation}. The distance from generator $p$ to GAP $i$ is $d_{pi}$, and the maximum distance between any generator in $P$ and its assigned GAP is $D$.
\item A set $H$ of waste fractions. Each dwelling $p$ has a generation rate $b_{ph}$ of waste fraction $h$.
\item A set $Y$ of collection frequencies. Each frequency $y$ has a parameter $a_y$ that indicates the number of days between two consecutive visits of the collection vehicle. For example, assume that the patterns considered are to collect the waste daily ($y_{1}$) or every two days ($y_{2}$) (i.e., $Y = \{y_{1}, y_{2}\}$). Thus, $a_{y_{1}} = 1$ and $a_{y_{2}} = 2$.
\end{itemize}

The proposed model is described in Eqs.~\eqref{eq:1}-\eqref{eq:13} where variable $t_{jhi}$ represents the number of bins of type $j$ and waste fraction $h$ installed in GAP $i$, variable $x_{pi}$ is 1 if generator $p$ is assigned to GAP $i$ and 0 otherwise, and variable $f_{hiy}$ is 1 if the waste fraction $h$ in GAP $i$ is collected with a frequency $y$, 0 otherwise.

Three objective functions are considered. Eq.~\eqref{eq:1} expresses the average collection frequency of the number of GAPs that are opened. The frequencies are divided by the number of days between two consecutive visits ($a_{y}$). \green{Therefore, for minimizing Eq.~\eqref{eq:1} the model aims both to use the smaller number of GAPs (an unused GAP respects $f_{hiy} = 0\ \forall\  y\in{Y}$) and, for those GAPs that are used, to impose a low collection frequency (associate with a large $a_{y}$). This aims at bounding the posterior routing costs of waste collection: either a GAP does not have to be visited because it remains unused or it has to be visited with a low frequency.} Eq.~\eqref{eq:2} represents the average distance between generators and the assigned GAPs. Eq.~\eqref{eq:3} represents the investment cost. 

\begin{align}
\label{eq:1}
\min &\frac{\sum_{\substack{h\in H,\ i\in I \\ y\in Y}}\left(\frac{f_{hiy}}{a_y}\right)}{\left\vert{}I\right\vert{} \left\vert{}H\right\vert{}} \\
\label{eq:2}
\min &\sum_{\substack{p \in P,\ i \in I}}\frac{\left(d_{pi} x_{pi}\right)}{\left\vert{}P\right\vert{}} &\\
\label{eq:3}
\min &\sum_{\substack{j \in J,\ h \in H \\ i \in I}}\left(t_{jhi} {cin}_{j}\right) & \\
\nonumber
\text{Subject to} & & \\
\label{eq:4}
\sum_{i\in{}I}\left(x_{pi}\right) ={} & 1, \ \forall\ p\in P & \\
\label{eq:5}
\sum_{{j \in J,\ h \in H}}\left(t_{jhi} e_j\right) \leq{} & {Es}_i, \ \forall\ i\in I & \\
\label{eq:6}
\sum_{p\in{P},\ y\in{Y}}\left(b_{ph} x_{pi} f_{hiy} a_y\right) \leq{} & \sum_{j\in{}J}\left({cap}_j t_{jhi}\right), \ \forall\ h \in H,\ i \in I & \\
\label{eq:7}
\sum_{y \in Y} f_{hiy} \leq{} & 1, \ \forall\ h\in H,\ i\in I \\
\label{eq:8}
|P|\sum_{y\in Y} f_{hiy} \geq{} & \green{\sum_{p\in P} x_{pi}}, \ \forall\ h\in H,\ i\in I & \\
\label{eq:10}
d_{pi} x_{pi}\leq{} & D, \ \forall\ p\in P,\ i\in I & \\
\label{eq:11}
x_{pi}&\in{}\left[0,1\right], \forall\ p\in P,\ i\in I &\\
\label{eq:12}
f_{hiy}&\in{}\left[0,1\right], \forall\ h\in H,\ i\in I,\ y\in Y &\\
\label{eq:13}
t_{jhi}&\in{{\mathbb{Z}}_{0}^{+}}, \forall\ j\in J,\ h\in H,\ i\in I&
\end{align}

\green{Nine sets of constraints} are included in the problem formulation. Eq.~\eqref{eq:4} enforces that every dwelling is assigned to one GAP. Eq.~\eqref{eq:5} controls that the maximum available space in each GAP is not exceeded. Eq.~\eqref{eq:6} ensures that the volume assigned to one GAP is not larger than the storage capacity installed in that GAP. Eq.~\eqref{eq:7} establishes that only one collection frequency pattern is assigned to each waste fraction of each GAP. Eq.~\eqref{eq:8} ensures that if (at least) one dwelling is assigned to a GAP, a frequency pattern to collect that garbage is selected. Eq.~\eqref{eq:10} establishes the maximum threshold distance between a dwelling and its assigned GAP. Eqs.~\eqref{eq:11} and \eqref{eq:12} indicate the binary nature of the variables. Finally, Eq.~\eqref{eq:13} states that \green{variables $t_{jhi}$ are} non-negative integer.

The proposed mathematical formulation (Eqs.~\eqref{eq:1}-\eqref{eq:13}) is not linear due to the presence of Eq.~\eqref{eq:6}. \magenta{In the literature, non-linear integer problems are often transformed through linearization techniques since linear formulations can benefit from a larger variety of resolution methods~\citep{glover1974converting}. However, when linearizing the formulation it is important to not increase the number of integer variables (which are closely related to the computational complexity of the problem). Thus, in this article the linearization technique proposed by~\citet{glover1984improved}, which respects this rule, is applied.} Eq.~\eqref{eq:6} is replaced with Eqs.~\eqref{eq:14}-\eqref{eq:18} through the introduction of continuous variable $u_{phiy}$. Finally, the linear equivalent formulation of the model is composed by Eqs.~\eqref{eq:1}-\eqref{eq:5} and~\eqref{eq:7}-\eqref{eq:18}.

\begin{align}
\label{eq:14}
\sum_{\substack{p \in P,\ y \in Y}}\left[b_{ph} a_y (u_{phiy} + f_{hiy} - 1 + x_{p,i})\right] & \leq{} \sum_{j \in J}\left({cap}_j t_{jhi}\right),\ \forall\ h \in H,\ i \in I
\end{align}
\begin{align}
\label{eq:15}
u_{phiy} & \geq{} 1 - x_{p,i} - f_{hiy},\ \forall\ p \in P,\ h \in H,\ i \in I,\ y \in Y\\
\label{eq:16}
u_{phiy} & \leq{} 1 - f_{hiy},\ \forall\ p \in P,\ h \in H,\ i \in I,\ y \in Y\\
\label{eq:17}
u_{phiy} & \leq{} 1 - x_{pi},\ \forall\ p \in P,\ h \in H,\ i \in I,\ y \in Y \\
\label{eq:18}
u_{phiy} & \geq{} 0,\ \forall\ p \in P,\ h \in H,\ i \in I,\ y \in Y
\end{align}

\subsection{Solution approach}
\label{subsec:solution_approach}

The proposed approach it based on a variation of the $\varepsilon$-constraint method, initially proposed by~\citet{haimesX1971bicriterion}. \green{Although $\varepsilon$-constraint has overcome some of the main disadvantages of the weighted sum, it has also some important drawbacks, e.g., the generation of weakly efficient solutions~\citep{mavrotas2009effective}.} Therefore, novel variants of this method have emerged to improve the original version, such as the one used in this paper: augmented $\varepsilon$-constraint which was proposed by~\citet{mavrotas2009effective} (AUGMECON) and later improved by~\citet{mavrotas2013improved} (AUGMECON2).

In order to apply the AUGMECON2 method, the range of the objectives function over the efficient set of solutions, i.e., the interval between the best (ideal) and the worst (nadir) values that the objective can assume within the Pareto front, are needed. These values are used as the limits of the interval within which the parameters $\varepsilon$ associated to each criteria can vary. Although AUGMECON2 should be able to deal with approximated ranges that are larger than the range over the efficient set, using the actual efficient values (or at least close approximations) can reduce the computing time~\citep{mavrotas2013improved}. This is particularly important in problems that are hard to solve, as is the case of the NP-hard problem addressed in this paper. Unlike the ideal value that can be closely approximated through single objective optimization, estimating the nadir value over the Pareto front is generally much harder~\citep{ehrgott2003computation}. Since the target is the efficient nadir value (within the Pareto front), approaches entirely based on individual optimization are not very convenient. For example, \citet{ehrgott2002constructing} estimated the nadir values from a payoff table composed by the results of the individual optimization of each objective. However, this payoff table can have Pareto suboptimal solutions in case these single objective problems have alternative optima~\citep{reeves1988minimum}. Another example are~\citet{zhang2014simple}, who calculate the nadir values from the optimization of the corresponding inverse objective, which is not \green{guaranteed} to be the efficient nadir value.

In the original article that introduced AUGMECON, \citet{mavrotas2009effective} applied lexicographic optimization. In an initial stage, they optimize the first criteria as a single objective problem. In a second stage, they optimize the second criteria in a single objective fashion but including a restriction that prevents the first criteria to obtain a worse value than the one of the initial stage. This continues until the last criteria is optimized, generating a non-dominated solution. Moreover, the authors distributed a ready-made version in GAMS for the sake of reproduction. However, when tested on the model of the previous Section, this implementation was not able to converge to a feasible solution whithin reasonable execution times. This might be related with the NP-hard nature of the addressed problem, as happened with a similar biobjective problem in~\citet{rossit2018municipal}. In this article, the use of warm starts is proposed for enhancing the convergence of the lexicographic approach~\citep{ralphs2006duality}. Since, in the lexicographic approach the solution of the previous stage constitutes a feasible solution of the following stage, this solution can be used for initializing the second stage. Providing an incumbent solution from the beginning, reduces the size of the branch-and-cut tree and allows the implementation of heuristics that require an incumbent solution~\citep{cplex2015v12}. 

In another case, \citet{tralhaoX2010multiobjective} obtained the best values with an unbalanced weighted sum method. When they optimize one of the objectives, they still assign very small weights to the other "[...] to ensure the identification of a non-dominated solution."~\citep[][pag. 2423]{tralhaoX2010multiobjective}.
However, in their approach weights are not normalized which can affect the result. If the objectives have different units of measurement or different ranges the weighted sum can bias the results towards those attributes with higher absolute values~\citep[see][]{ballestero2013multiple}. One reason for not normalizing, is that there is not prior information about the magnitudes of the objectives. To overcome this issue, in our previous work~\citep{rossit2018municipal} the results of single objective optimization are used for the normalization, as was proposed in~\citet{rossit2018thesis}. The method works as follows. Let $K$ be the set of the optimization criteria of a minimization multiobjective problem and $w_{k}$ be the weight assigned to criteria $k$ defined by function $Obj_{k}$. Moreover, $Obj_{k}^{w}$ and $Obj_{k}^{b}$ are the worst and the best value of objective $k$ taken from single objective optimization. Eq.~\eqref{eq:ourNormalization} is applied with the aim of finding the efficient range of the objectives individual optima of the objective $k'\in K$, with $w_{k'}~\gg~w_{k}\ \forall\ k\in~K,\ k\neq~k'$.

\begin{equation}
\label{eq:ourNormalization}
\min \left[w_{k'} \frac{Obj_{k'} - {Obj_{k'}^{b}}}{Obj^{w}_{k'} - Obj_{k'}^{b}} + \sum_{\substack{k\in{}K \\ k\neq{}k'}} \left(w_{k} \frac{Obj_{k} - {Obj_{k}^{b}}}{Obj_{k}^{w} - Obj_{k}^{b}}\right)\right]
\end{equation}

\section{A Pagerank-based heuristic for the GAP location problem}
\label{sec:heuristic}

The GAP location problem, as a variation of the CFLP, is NP-hard. Thus, competitive heuristics approaches are applied to generate feasible solutions in reasonable computing times~\citep{nesmachnow2014overview}. An idea applied in previous studies~\citep{brahimX2014roadside,massobrioX2017infrastructure,toutouh2018intelligence} is to define the problem over a weighted graph, in which the vertices represent the possible locations and the links give the weight/influence/importance of the linked vertices. 
Then, a method to sort the vertices according to the \textit{relevance} of each vertex on the whole system is applied, returning them in a vector. Finally, a constructive heuristic is used to configure the whole system iterating over the vector of sorted vertex. In literature, vertices have been sorted according to their PageRank value computed by using weighted PageRank algorithm \citep{massobrioX2017infrastructure}. PageRank can be defined as a voting algorithm. It was developed to compute the relevance of web pages in Internet taking into account the inbound and outbound links~\citep{Langville2011}. The key idea behind PageRank is to allow propagation of influence along the whole network of web pages, instead of just counting the number of other web pages pointing at the web page. The weighted PageRank is applied to a given directed graph $G=(V,E)$ defined by $V$ (a set of vertices) and $E$ (a set of edges). The algorithm starts by initializing the PageRank value of each vertex $v_i$ to a fixed value $d$, i.e., $PR^W(v_i)=d,\  \forall v_i \in V$. $d$ is known as the \textit{dumping parameter} and its default value is 0.85. Then, an iterative process is performed until a stop condition is reached (the convergence value is below a given threshold or a maximum number of iterations \green{is} performed). During this iterative process, $PR^W(v_i)$ is computed according to Equation~\eqref{eq:pagerank}. In that equation, $In(v_i)$ is the set of vertices that point to it (\textit{predecessors}), and $Out(v_i)$ is the set of vertices that $v_i$ points to (\textit{successors}), and $w_{ij}$ is \green{the weight} for the edge that connects $v_i$ and $v_j$. 

\begin{equation}
\centering
\small
\label{eq:pagerank}
PR^W(v_i) = (1-d) + d \times \left( \sum_{v_j \in In(v_i)} w_{ij} \times \frac{PR^W(v_j)}{ \sum\limits_{v_k \in Out(v_j)} w_{jk}}\right)
\end{equation}

For applying PageRank to the GAP location problem, the city is modeled as a fully connected weighted graph $G=(V,E)$, 
by taking into account topological information (i.e., streets, collection points, and generator). 
$G$ is defined by the set of location of collection points $P$ and the set of edges $E$. The weight of each edge $w_{j,k}$ is given by the weight of the arc between two collection points, which is computed according to $w_{jk} = \frac{b_j + b_k}{d_{j,k}}$, which relates the impact of the waste generated in the generator in such collection points and their distance.
Thus, the tentative locations of collection points are ranked in a sorted vector $I^{PR}$ in which $i^{PR}_j, i^{PR}_k \in I$, $j<k \Leftrightarrow PR^W(i^{PR}_j) > PR^W(s^{PR}_k$). 
Once the collection points are sorted in $I^{PR}$, a constructive heuristic is applied to select a collection point configuration and locate it. Considering the particular structure of PageRank algorithms, the optimization criteria considered are slightly different from the ones proposed in Section~\ref{sec:exact_multiobjective}. This heuristic considers that the collection frequency is set to one day for all the GAPs where a waste bin is installed, i.e., once a day the collection points are emptied. Then, additionally to the objectives of minimizing the total average distance between the \green{generator} and the assigned GAPs and minimizing the number of bins installed, it aims to maximize the total amount of waste collected. This last objective is valuable in developing countries where generally the formal MSW system has to compete with informal collection \citep{rathoreX2019location,steuerX2017analysis}. Garbage source classification is not considered in this approach.

PageRank operates in two steps.  First, the \textit{sorting step}, in which, it sorts the possible location of the collection points by using PageRank to obtain $I^{PR}$. Second, the \textit{constructive step}, in which, it iterates over $I^{PR}$ to select the best configuration of the containers for such a collection point according to one of the three objectives of the problem. Depending \green{on} which objective is pursued three different variants of the heuristic are developed:
\begin{itemize}
\item \textit{Pagerank-Vol}: selects the configuration that collects the maximum volume of waste from the nearby generator. If more than one containers configuration collect the same maximum of waste, the one with the cheapest installation cost is selected.
\item \textit{Pagerank-Dist}: considers the solutions that collects at least all the generated waste by the nearest dwelling, then it selects the one that allows users to walk the shortest distance. If more than one configuration have the same minimum distance, the one that collects the maximum volume of waste is selected.
\item \textit{Pagerank-Cost}: evaluates the solutions that collects at least all the generated waste by the nearest generator, then it considers the one with the cheapest installation costs. If more than one configuration have the same minimum cost, the one that collects the maximum volume of waste is selected.
\end{itemize}

The general skeleton of the PageRank heuristic is described in Algorithm~\ref{alg:PR}.
\begin{algorithm}
\caption{Skeleton of \textit{Pagerank-Vol}, \textit{Pagerank-Dist}, and \textit{Pagerank-Cost}}\label{alg:PR}
\begin{algorithmic}[1]
\Procedure{Pagerank}{$I,P,J,H$}
\State $G \gets getGraph(I,P,H)$ \Comment{Get graph G=(V,E)}
\State $I^{PR} \gets PR^W(G)$\Comment{Sort step: Obtain $I^{PR}$}
\State $Solution \gets newVector(length(I^{PR}))$
\State $i \gets 0$

\Comment{Loop while there are location points not configured}

\Comment{and waste to be collected}
\While{$i<length(I^{PR})$ \textbf{and} $\sum_{\substack{p\in{}P \\ h\in{}H}}{\left(b_{ph}\right)} > 0 $}

\Comment{Constructive step}
\State $ Solution[I^{PR}(i)] \gets selectTheBestChoice(I^{PR}(i),P,J,H)$
\State $i \gets i + 1$
\EndWhile\label{euclidendwhile}
\State \textbf{return} $Solution$
\EndProcedure
\end{algorithmic}
\end{algorithm}

The model solved with PageRank algorithms considers that every opened GAP is emptied daily. Moreover, it does not require to collect all the garbage (although this quantity is maximize as one of the objectives). Both situations reduce the minimal required installed capacity in comparison to the model solved in Section~\ref{sec:exact_multiobjective}. This way these solutions can provide a lower bound in terms of installment cost to the solutions provided by the exact approach.

\FloatBarrier

\section{Experimental evaluation}
\label{sec:experimental_evaluation}

This section describes the experimental evaluation of the proposed computational methods for solving the GAPs location problem. \green{The experimentation was performed on a Core~i7 processor, with 16 GB of RAM memory. The problem was modeled in C++ and the resolution of the the exact approach was performed with the parallel mode of CPLEX 12.7.1.}. Section \ref{subsec:description_scenarios} describes the real problem scenarios considered in the study. The experimental evaluation of the exact approach is reported in Section~\ref{subsec:resultsExact} and the evaluation of the PageRank heuristic approach is reported in Section~\ref{subsec:resultsHeuristic}. Finally, in Section~\ref{subsec:analysis_results} the analysis of the results is performed.

\subsection{Real scenarios for the GAPs location problem}
\label{subsec:description_scenarios}

The proposed resolution approaches were evaluated on real scenarios built using information from the cities of Montevideo, Uruguay, and Bah\'{i}a Blanca, Argentina. Montevideo is the capital city and the largest urban agglomeration of Uruguay, with a population of about 1.5 million people. \blue{Nowadays, Montevideo has a community bins system in the highly populated areas which supports source classification. Waste is classified in two main fractions: recyclable waste and 'mixed' waste~\green{\citep{manualresiduos}}, which includes all the types of waste that can not be recycled in the city. Due to variations of public policies and the incorporation of new technologies, the City Hall can increase the lists of materials that are considered recyclable waste along the time~\green{\citep{manualresiduos}}. For this article, recyclable waste is composed mainly by paper, plastic, metal, and glass whereas mixed waste is composed mainly by organic waste and other humid domestic waste (e.g., diapers and hand towels), ceramic, sweepings, bones, among others.} The model is applied to three different scenarios of Montevideo (Table~\ref{tab:scenarios}). Considering information about bins alread used in the city, from the government of Montevideo~\green{\citep{MML}}, three bin types ($j_1$, $j_2$, and $j_3$) are considered, according to the normal utilization in Montevideo. The values of parameters $c_j$, $C_j$, and $e_j$ are: 1000 monetary units (m.u.), 1 $m^{3}$ and 1 $m^{2}$ for bin type $j_1$, 2000 m.u., 2 $m^{3}$ and 2 $m^{2}$ for bin type $j_2$, and 3000 m.u., 3 $m^{3}$ and 3 $m^{2}$ for bin type $j_3$.

Bah\'{i}a Blanca is one of the main export ports and industrial centers of Argentina and, with approximately 300,000 inhabitants, is the largest city in the South of Argentina. The MSW collection system is still based on an unsorted waste door-to-door system but the local government aims to use a community bins system instead which has been proved to be a valid strategy to reduce the high expenses of the system~\citep{bonomoX2012method} specially the logistic costs which are remarkably high in Argentina~\citep{broz2018argentinian}. Three scenarios from Bah\'{i}a Blanca are considered (Table~\ref{tab:scenarios}). Since this city has not implemented a community bins system yet, three types of commercial bins are used for the test. The values of parameters $c_j$, $C_j$, and $e_j$ for each type are: 2120 monetary units (m.u.), 1.1 $m^{3}$ and 1.34 $m^{2}$ for bin type $j_1$, 3170 m.u., 1.73 $m^{3}$ and 1.67 $m^{2}$ for bin type $j_2$, and 5380 m.u., 3.1 $m^{3}$ and 2.5 $m^{2}$ for bin type $j_3$.

\begin{table}[h]
\caption{Details of the proposed scenarios for the GAPs location problem.\label{tab:scenarios}}
\renewcommand{\arraystretch}{0.80}
\setlength{\tabcolsep}{5pt}
{\footnotesize
\begin{tabular*}{\hsize}{c|lllrr}
\toprule
City & Id & Neighborhood & \magenta{Main features} & \begin{tabular}{r}Number of\\potential GAPs\end{tabular} & \begin{tabular}{r}Estimated\\population\end{tabular}  \\ 
\midrule
 \multirow{6}{*}{Montevideo}
 & MVD\_1 & Downtown & \begin{tabular}{l}\magenta{residential,}\\\magenta{commercial,}\\\magenta{administrative} \end{tabular} & 59 & 5211 \\
 \cline{2-6}
 & MVD\_2 & Punta Carretas & \begin{tabular}{l}\magenta{high-income}\\\magenta{residential,} \\\magenta{commercial} \end{tabular} & 63 & 7767 \\
 \cline{2-6}
 & MVD\_3 & Villa Espa\~{n}ola & \begin{tabular}{l} \magenta{medium-low-income}\\\magenta{residential} \end{tabular} &70 & 2528 \\  \midrule 
\multirow{6}{*}{Bah\'{i}a Blanca}
 & BBCA\_1 & Barrio Universitario & \begin{tabular}{l}\magenta{residential,}\\\magenta{commercial,}\\\magenta{administrative} \end{tabular} & 88 & 7903 \\
 \cline{2-6}
 & BBCA\_2 & La Falda & \begin{tabular}{l}\magenta{middle-low-income}\\\magenta{residential} \end{tabular} & 99 & 4929 \\
 \cline{2-6}
 & BBCA\_3 & Villa Mitre & \begin{tabular}{l}\magenta{middle-income}\\\magenta{residential,} \\\magenta{commercial} \end{tabular} & 115 & 3033  \\
 \bottomrule
 \end{tabular*}
}
\end{table}

\magenta{The proposed study is focused on residential areas, mainly because in the studied cities there is no clear difference between administrative, commercial, and residential areas (i.e., most of administration offices and shops are widely distributed over many neighborhoods). Therefore, the set of scenarios of each city was chosen with the aim of considering representative residential neighborhoods. In the case of Montevideo, MVD\_1 is a central residential area of the city, which also has several administrative buildings and a relevant commercial activity. MVD\_2 is high-income residential zone with an important commercial activity. Finally, MVD\_3 is a residential lower-middle-income neighborhood, with fewer commercial buildings. In the case of Bah\'ia Blanca, BBCA\_1 also has the three aforementioned characteristics, being a residential area that also includes several administrative and commercial activities. BBCA\_3 is a middle-income residential area with an important commercial activity. Finally, BBCA\_2 is mainly a lower-middle-income residential neighborhood.}

The population density for Montevideo (per square block) was retrieved from the City Hall website~\green{\citep{sigMontevideo}}, while the garbage generation rate and composition was retrieved from a study carried out by the national government~\citep{planmaestro2012}. For Bah\'{i}a Blanca scenarios, the population distribution was obtained from the local government~\citep{direccion2010}, the composition of waste was retrieved from a report performed by a local research center~\citep{plapiqui2013} and the density of the different garbage flows from a characterization study performed in Argentina~\citep{pettigianiX2013caracterizacion}. Distances between generators and GAPs were calculated using a modified version of the R package \textit{osmar}~\citep{eugster2013osmar} in order to obtain walking (undirected) distances. As in many works done on the field \citep{coutinhoX2012bi,tralhaoX2010multiobjective}, the generators are aggregated in order to maintain the tractability of the problem. In our case, generators were aggregated in linear sectors. The values of the other parameters for all the studied scenarios are: $D = 300m$ and $Es_{i} = 5m^2$ for every GAP $i$. Finally, three frequencies are considered: $y_1$, $y_2$, and $y_3$ with $a_{y}$ equal to 1, 2, and 3, respectively. \magenta{Daily collection is effective to prevent environmental problems, but it can yield sub-optimal results, mainly due to operational costs when bins are not full everyday. Although a frequency of three days is not common, it was included in the model since it is currently implemented in scenarios MVD\_2 and MVD\_3 by the administration of Montevideo. This set of frequencies respects the highest and lowest collection frequencies that are currently implemented in Montevideo, $y_1$ for scenario MVD\_1 and $y_3$ for scenarios MVD\_2 and MVD\_3.}

\FloatBarrier

\subsection{Experimental evaluation: exact approach}
\label{subsec:resultsExact}

This section reports the results of the exact approach for solving the GAP location problem. A two-phase procedure is applied for obtaining the solutions. The first phase consists on approximating the efficient range \green{of} the objectives and the second phase consists on the application of AUGMECON2 to obtain multiobjective solutions. For the sake of brevity, only for the scenarios with unclassified waste the two stages are presented. \green{For the results of first phase of scenarios with source classified waste refer to~\ref{appendix:first_stage}}.

As stated, in the first phase four different methods are used to approximate the ranges of the objectives over the efficient set. These are: single objective optimization, (unbalanced) weighted sum, and the two lexicographic approaches: straightforward lexicographic optimization and the lexicographic method that uses warm starts. The time limit of the runs was set to 4200 seconds. For the two lexicographic approaches, this time limit was imposed on each stage of the optimization stages, i.e., when each criteria was optimized. The results are presented in Tables~\ref{tab:obj_range_MVDI_and_MVDII_unclas}-\ref{tab:obj_range_BBCAII_and_BBCAIII_unclas}. For each scenario, the ideal and nadir values, i.e., $Ideal_k$ and $Nadir_k$, within the approximated efficient set (non-dominated solutions) are selected for each objective $k \in K$. Then, the relative deviation of the value of each objective ($\Delta Obj_k$) was computed by Eq.~\ref{eq:percentage}.  

\begin{equation} \label{eq:percentage}
 \Delta Obj_k = \frac{Value - Ideal_k}{Nadir_k - Ideal_k} \cdot 100\%
\end{equation}

\magenta{Finally, the Euclidean norm is applied to obtain an overall estimation of the deviation of each solution using the formula $L^2 = {\left( \sum_{k \in K} {\left( {\Delta Obj_k}_{k} \right)}^{2} \right)}^{1/2}$. This norm, which has been used in similar works~\citep{coutinhoX2012bi,tralhaoX2010multiobjective}, measures the Euclidean distance to the ideal multiobjective solution giving to decision-makers an estimation of how far a solution is from the ideal (infeasible) solution~\citep{deb2001multi}.}

The results reported in Tables~\ref{tab:obj_range_MVDI_and_MVDII_unclas}--\ref{tab:obj_range_BBCAII_and_BBCAIII_unclas} 
are organized, from left to right, in the following columns: the method used to approximate the efficient ranges, the optimized objective, the obtained value for each of the three individual objectives and its individual deviation, the overall deviation $L^2$, the execution time (in seconds), and either if the solution is dominated or not by any other solution of the four methods. \magenta{The execution time is a relevant feature considering the aforementioned aim of this work to provide efficient and flexible methods to decision-makers that allow them to explore different compromising solutions. Moreover, efficient methods that provide relatively good solutions with few computational resources allow to resolve the GAP location problem when the input parameters vary significantly, e.g., when there are important variations of the population density or when an street is blocked to vehicular access.}
In the case of lexicographic optimization, the column of \textit{optimized objective} indicates the order in which the objectives were optimized. For example, if the \textit{optimized objective} is $Obj_1, Obj_2$ it means that $Obj_1$ was optimized on the first stage, then $Obj_2$ on the second stage and finally the remaining objective $Obj_3$ on the third stage. It is considered that lexicographic optimization reaches a feasible solution only if the solver is able to find a feasible solution in the third stage. The first stage of lexicographic optimization is the single objective optimization of the first objective that is optimized. Similarly, the unbalanced weighted sum approach requires the initial objective ranges from single objective optimization to apply Eq.~\eqref{eq:ourNormalization}. For the analysis reported in this article, the results of the single objective optimization are considered as an input of the weighted sum and the lexicographic methods; therefore, the execution times of these methods do not include the time required to perform the single objective optimization.

\begin{sidewaystable}
\centering
\setlength{\tabcolsep}{4pt}
\renewcommand{\arraystretch}{0.7}
\caption{Objectives range approximation for scenarios of MVD\_1 and MVD\_2 with unclassified waste.\label{tab:obj_range_MVDI_and_MVDII_unclas}}
{\scriptsize 
\begin{tabular}{cc rrr rrr rr c}
\toprule
Method & \begin{tabular}{c} Optimized\\objective \end{tabular} & $Obj_1$ & $\Delta Obj_1$ (\%) & $Obj_2$ & $\Delta Obj_1$ (\%) & $Obj_3$ & $\Delta Obj_1$ (\%) & $L_2$ (\%)  & \begin{tabular}{r}Execution\\time (s)\end{tabular}  & Dominance \\ \midrule
\multicolumn{11}{c}{MVD\_1} \\ \midrule
\multirow{3}{2.1cm}{Single objective optimization} 
 & $Obj_1$ & 0.1610 & 0.00\%  & 193.49 & 169.43\% & 29500 & 209.32\% & 269.30\% & 4207.30 & D \\*
 & $Obj_2$ & 0.5989 & 65.96\% & 0.00   & 0.00\%   & 29500 & 209.32\% & 219.47\% & 0.55 & D \\*
 & $Obj_3$ & 0.3079 & 22.13\% & 184.92 & 161.93\% & 4800  & 0.00\%   & 163.43\% & 4229.23 & D \\ \midrule
 \multirow{3}{2cm}{Weighted sum} 
  & $Obj_1$ & 0.1638 & 0.43\%  & 117.24 & 102.66\% & 7500  & 22.88\% & 105.18\% & 4200.48 & D \\
 & $Obj_2$ & 0.3531 & 28.94\% & 0.00   & 0.00\%   & 16200 & 96.61\% & 100.85\% & 3.53  & ND  \\
 & $Obj_3$ & 0.2034 & 6.38\%  & 114.12 & 99.93\%  & 4800  & 0.00\%  & 100.13\% & 4347.67 & D \\
  \midrule
  \multirow{6}{2cm}{Lexicographic} 
 & $Obj_1, Obj_2$ & \multicolumn{9}{c}{\textit{No feasible solution found}} \\*	
 & $Obj_1, Obj_3$ & \multicolumn{9}{c}{\textit{No feasible solution found}} \\*	
 & $Obj_2, Obj_1$ & 0.3475 & 28.09\%  & 0.00 & 0.00\% & 16600 & 100.00\% & 103.87\% & 0.12 & ND \\*
 & $Obj_2, Obj_3$ & 0.8249(*) & 100.00\% & 0.00 & 0.00\% & 7400  & 22.03\%  & 102.40\% & 0.12 & ND \\*
 & $Obj_3, Obj_1$ & \multicolumn{9}{c}{\textit{No feasible solution found}}  \\*
 & $Obj_3, Obj_2$ & \multicolumn{9}{c}{\textit{No feasible solution found}}  \\	\midrule	
\multirow{6}{2cm}{Lexicographic with warm starts} 
 & $Obj_1, Obj_2$ & 0.1610 & 0.00\%   & 109.47 & 95.86\%  & 7500  & 22.88\%  & 98.55\%  & 8428.92 & ND \\*
 & $Obj_1, Obj_3$ & 0.1610 & 0.00\%   & 114.20(*) & 100.00\% & 6000  & 10.17\%  & 100.52\% & 8400.39 & ND \\*
 & $Obj_2, Obj_1$ & 0.3475 & 28.09\%  & 0.00   & 0.00\%   & 16600 & 100.00\% & 103.87\% & 0.25  &  ND \\*
 & $Obj_2, Obj_3$ & 0.8249(*) & 100.00\% & 0.00   & 0.00\%   & 7400  & 22.03\%  & 102.40\% & 0.28  &  ND \\*
 & $Obj_3, Obj_1$ & 0.1695 & 1.28\%   & 111.12 & 97.30\%  & 4800  & 0.00\%   & 97.31\%  & 8401.80 & ND \\*
 & $Obj_3, Obj_2$ & 0.3729 & 31.91\%  & 67.95  & 59.50\%  & 4800  & 0.00\%   & 67.52\%  & 8406.37 & ND \\ \midrule
\multicolumn{11}{c}{MVD\_2} \\ 
\midrule
\multirow{3}{2.1cm}{Single objective optimization} 
 & $Obj_1$ & 0.1016 & 0.00\%  & 176.48 & 154.54\% & 32000 & 306.38\% & 343.15\% & 4200.13 & D \\
 & $Obj_2$ & 0.5964 & 97.94\% & 0.00   & 0.00\%   & 32000 & 306.38\% & 321.66\% & 0.61 & D \\
 & $Obj_3$ & 0.4922 & 77.32\% & 185.70 & 162.61\% & 3300  & 1.06\%   & 180.06\% & 4282.4 & D \\
\midrule
 \multirow{3}{2cm}{Weighted sum} 
 & $Obj_1$ & 0.1016 & 0.00\%  & 94.36 & 82.63\% & 7500  & 45.74\%  & 94.44\% & 4222.24 & ND \\
 & $Obj_2$ & 0.3359 & 46.39\% & 0.00  & 0.00\%  & 12400 & 97.87\% & 108.31\% & 1.55  & D \\
 & $Obj_3$ & 0.1406 & 7.73\%  & 83.14 & 72.80\% & 3200  & 0.00\%   & 73.21\%  & 4201.73 & ND \\
  \midrule
  \multirow{6}{2cm}{Lexicographic} 
 & $Obj_1, Obj_2$ & \multicolumn{9}{c}{\textit{No feasible solution found}} \\*	
 & $Obj_1, Obj_3$ & 0.1016 & 0.00\%   & 117.28 & 102.70\% & 3400  & 2.13\%   & 102.72\% & 8400.16 & D \\
 & $Obj_2, Obj_1$ & 0.3333 & 45.88\%  & 0.00   & 0.00\%   & 12600 & 100.00\% & 110.02\% & 0.18  & ND \\
 & $Obj_2, Obj_3$ & 0.6068(*) & 100.00\% & 0.00   & 0.00\%   & 7200  & 42.55\%  & 108.68\% & 0.12  & ND \\
 & $Obj_3, Obj_1$ & \multicolumn{9}{c}{\textit{No feasible solution found}}  \\*
 & $Obj_3, Obj_2$ & \multicolumn{9}{c}{\textit{No feasible solution found}}  \\	\midrule	
\multirow{6}{2cm}{Lexicographic with warm starts} 
  & $Obj_1, Obj_2$ & 0.1016 & 0.00\%   & 51.75  & 45.32\% & 8500  & 56.38\% & 72.34\% & 8432.20 & ND \\
 & $Obj_1, Obj_3$ & 0.1016 & 0.00\%   & 113.59 & 99.47\% & 3400  & 2.13\%   & 99.49\%  & 8406.13 & ND \\
 & $Obj_2, Obj_1$ & 0.3333 & 45.88\%  & 0.00   & 0.00\%  & 12600 & 100.00\% & 110.02\% & 0.38    & ND \\
 & $Obj_2, Obj_3$ & 0.6068(*) & 100.00\% & 0.00   & 0.00\%  & 7200  & 42.55\%  & 108.68\% & 0.34   & ND \\
 & $Obj_3, Obj_1$ & 0.1094 & 1.55\%   & 80.94  & 70.88\% & 3300  & 1.06\%   & 70.90\%  & 8402.61 & ND \\
 & $Obj_3, Obj_2$ & 0.3438 & 47.94\%  & 53.06  & 46.46\% & 3300  & 1.06\%   & 66.77\%  & 8404.97 & ND \\
\bottomrule
\end{tabular}
}
\end{sidewaystable}

\begin{sidewaystable}
\centering
\setlength{\tabcolsep}{4pt}
\renewcommand{\arraystretch}{0.7}
{\scriptsize 
\caption{Objectives range approximation for scenarios of MVD\_3 and BBCA\_1 with unclassified waste.\label{tab:obj_range_MVDIII_and_BBCAI_unclas}}
\begin{tabular}{cc| rrrrrrrrc}
\toprule
Method & \begin{tabular}{c} Optimized\\objective \end{tabular} & $Obj_1$ & $\Delta Obj_1$ (\%) & $Obj_2$ & $\Delta Obj_1$ (\%) & $Obj_3$ & $\Delta Obj_1$ (\%) & $L_2$ (\%)  & \begin{tabular}{r}Execution\\time (s)\end{tabular}  & Dominance \\ \midrule
\multicolumn{11}{c}{MVD\_3} \\ \midrule
\multirow{3}{2.1cm}{Single objective optimization} 
 & $Obj_1$ & 0.1016 & 0.00\%  & 176.48 & 154.54\% & 32000 & 543.40\% & 564.94\% & 4200.13 & D \\
 & $Obj_2$ & 0.5964 & 97.94\% & 0.00   & 0.00\%   & 32000 & 543.40\% & 552.15\% & 0.61 & D \\
 & $Obj_3$ & 0.4922 & 77.32\% & 185.70 & 162.61\% & 3300  & 1.89\%   & 180.07\% & 4282.42 & D \\ \midrule
 \multirow{3}{2cm}{Weighted sum} 
 & $Obj_1$ & 0.1016 & 0.00\%  & 94.36 & 82.63\% & 7500  & 81.13\%  & 115.80\% & 4222.24 & ND \\
 & $Obj_2$ & 0.3359 & 46.39\% & 0.00  & 0.00\%  & 12400 & 173.58\% & 179.68\% & 1.55 & D \\
 & $Obj_3$ & 0.1406 & 7.73\%  & 83.14 & 72.80\% & 3200  & 0.00\%   & 73.21\%  & 4201.73 & ND \\
  \midrule
  \multirow{6}{2cm}{Lexicographic} 
 & $Obj_1, Obj_2$ & \multicolumn{9}{c}{\textit{No feasible solution found}} \\*	
 & $Obj_1, Obj_3$ & \multicolumn{9}{c}{\textit{No feasible solution found}} \\*	
 & $Obj_2, Obj_1$ & 0.3475 & 28.09\%  & 0.00 & 0.00\% & 16600 & 100.00\% & 103.87\% & 0.12 & ND \\*
 & $Obj_2, Obj_3$ & 0.8249(*) & 100.00\% & 0.00 & 0.00\% & 7400  & 22.03\%  & 102.40\% & 0.12 & ND \\*
 & $Obj_3, Obj_1$ & \multicolumn{9}{c}{\textit{No feasible solution found}}  \\*
 & $Obj_3, Obj_2$ & \multicolumn{9}{c}{\textit{No feasible solution found}}  \\	\midrule	
\multirow{6}{2cm}{Lexicographic with warm starts} 
 & $Obj_1, Obj_2$ & 0.1610 & 0.00\%   & 109.47 & 95.86\%  & 7500  & 22.88\%  & 98.55\%  & 8428.92 & ND \\*
 & $Obj_1, Obj_3$ & 0.1610 & 0.00\%   & 114.20(*) & 100.00\% & 6000  & 10.17\%  & 100.52\% & 8400.39 & ND \\*
 & $Obj_2, Obj_1$ & 0.3475 & 28.09\%  & 0.00   & 0.00\%   & 16600 & 100.00\% & 103.87\% & 0.25  &  ND \\*
 & $Obj_2, Obj_3$ & 0.8249(*) & 100.00\% & 0.00   & 0.00\%   & 7400  & 22.03\%  & 102.40\% & 0.28  &  ND \\*
 & $Obj_3, Obj_1$ & 0.1695 & 1.28\%   & 111.12 & 97.30\%  & 4800  & 0.00\%   & 97.31\%  & 8401.80 & ND \\*
 & $Obj_3, Obj_2$ & 0.3729 & 31.91\%  & 67.95  & 59.50\%  & 4800  & 0.00\%   & 67.52\%  & 8406.37 & ND \\ \midrule
\multicolumn{11}{c}{BBCA\_1} \\ 
 \midrule
\multirow{3}{2.1cm}{Single objective optimization} 
 & $Obj_1$ & 0.0852 & 0.00\%   & 192.33 & 123.08\% & 714.6  & 285.57\% & 310.97\% & 4228 & D \\
 & $Obj_2$ & 0.6269 & 97.28\%  & 0.00   & 0.00\%   & 364.73 & 127.05\% & 160.02\% & 0.47 & D \\
 & $Obj_3$ & 0.6477 & 101.02\% & 180.30 & 115.38\% & 84.85  & 0.24\%  & 153.36\% & 4209.39 & D \\
\midrule
 \multirow{3}{2cm}{Weighted sum} 
 & $Obj_1$ & 0.0852 & 0.00\%  & 192.33 & 123.08\% & 714.6  & 285.57\% & 310.97\% & 4239.05 & D \\
 & $Obj_2$ & 0.6269 & 97.28\% & 0.00   & 0.00\%   & 364.73 & 127.05\% & 160.02\% & 0.44 & D \\
 & $Obj_3$ & 0.4129 & 58.84\% & 178.56 & 114.27\% & 85.92  & 0.73\%  & 128.53\% & 4207.84 & D \\
  \midrule
  \multirow{6}{2cm}{Lexicographic} 
 & $Obj_1, Obj_2$ & \multicolumn{9}{c}{\textit{No feasible solution found}} \\*	
 & $Obj_1, Obj_3$ & \multicolumn{9}{c}{\textit{No feasible solution found}} \\*	
 & $Obj_2, Obj_1$ & 0.3333 & 44.56\%  & 0.00 & 0.00\% & 305.02 & 100.00\% & 109.48\% & 2.38 &  ND \\
 & $Obj_2, Obj_3$ & 0.6420 & 100.00\% & 0.00 & 0.00\% & 186.56 & 46.33\% & 110.21\% & 1.89 & ND \\
 & $Obj_3, Obj_1$ & \multicolumn{9}{c}{\textit{No feasible solution found}}  \\*
 & $Obj_3, Obj_2$ & \multicolumn{9}{c}{\textit{No feasible solution found}}  \\	\midrule	
\multirow{6}{2cm}{Lexicographic with warm starts} 
 & $Obj_1, Obj_2$ & 0.0852 & 0.00\%   & 107.59 & 68.85\%  & 225.96 & 64.18\% & 94.13\% & 8412.16 & ND \\
 & $Obj_1, Obj_3$ & 0.0852 & 0.00\%   & 156.26 & 100.00\% & 134.50  & 22.74\%  & 102.55\% & 8438.80 & ND \\
 & $Obj_2, Obj_1$ & 0.3333 & 44.56\%  & 0.00   & 0.00\%   & 305.02 & 100.00\% & 109.48\% & 0.42 & ND \\
 & $Obj_2, Obj_3$ & 0.6420 & 100.00\% & 0.00   & 0.00\%   & 186.56 & 46.33\% & 110.21\% & 0.40 & ND \\
 & $Obj_3, Obj_1$ & 0.1515 & 11.90\%  & 147.35 & 94.30\%  & 84.40   & 0.04\%  & 95.05\% & 8403.25 & ND \\
 & $Obj_3, Obj_2$ & 0.2614 & 31.63\%  & 108.19 & 69.24\%  & 84.31  & 0.00\%  & 76.12\%  & 8402.94 & ND \\
\bottomrule
\end{tabular}
}
\end{sidewaystable}

\begin{sidewaystable}
\centering
\setlength{\tabcolsep}{4pt}
\renewcommand{\arraystretch}{0.7}
{\scriptsize
\caption{Objectives range approximation for scenarios of BBCA\_2 and BBCA\_3 with unclassified waste.\label{tab:obj_range_BBCAII_and_BBCAIII_unclas}}
\begin{tabular}{c c| rrrrrrrrc}
\toprule
Method & \begin{tabular}{c} Optimized\\objective \end{tabular} & $Obj_1$ & $\Delta Obj_1$ (\%) & $Obj_2$ & $\Delta Obj_1$ (\%) & $Obj_3$ & $\Delta Obj_1$ (\%) & $L_2$ (\%)  & \begin{tabular}{r}Execution\\time (s)\end{tabular}  & Dominance \\ \midrule
\multicolumn{11}{c}{BBCA\_2} \\ \midrule
\multirow{3}{2.1cm}{Single objective optimization} 
  & $Obj_1$ & 0.0281 & 0.00\%   & 195.85 & 106.57\% & 532.86 & 341.02\% & 357.28\% & 4200.13 & D \\
 & $Obj_2$ & 0.6452 & 209.98\% & 0.00   & 0.00\%   & 310.66 & 189.16\% & 282.62\% & 0.61 & D \\
 & $Obj_3$ & 0.5759 & 186.40\% & 212.78 & 115.79\% & 33.88  & 0.00\%   & 219.43\% & 4282.42 & D \\ \midrule
 \multirow{3}{2cm}{Weighted sum} 
 & $Obj_1$ & 0.0281 & 0.00\%   & 195.85 & 106.57\% & 532.86 & 341.02\% & 357.28\% & 4200.13 & D \\
 & $Obj_2$ & 0.6452 & 209.98\% & 0.00   & 0.00\%   & 310.66 & 189.16\% & 282.62\% & 0.61    & D \\
 & $Obj_3$ & 0.5759 & 186.40\% & 212.78 & 115.79\% & 33.88  & 0.00\%   & 219.43\% & 4282.42 & ND \\
  \midrule
  \multirow{6}{2cm}{Lexicographic} 
 & $Obj_1, Obj_2$ & \multicolumn{9}{c}{\textit{No feasible solution found}} \\*	
 & $Obj_1, Obj_3$ & \multicolumn{9}{c}{\textit{No feasible solution found}} \\*	
 & $Obj_2, Obj_1$ & 0.3220 & 100.00\% & 0.00 & 0.00\% & 180.2 & 100.00\% & 141.42\% & 0.24 & ND \\
 & $Obj_2, Obj_3$ & 0.3220 & 100.00\% & 0.00 & 0.00\% & 180.2 & 100.00\% & 141.42\% & 0.22 & ND \\*
 & $Obj_3, Obj_1$ & \multicolumn{9}{c}{\textit{No feasible solution found}}  \\*
 & $Obj_3, Obj_2$ & \multicolumn{9}{c}{\textit{No feasible solution found}}  \\	\midrule	
\multirow{6}{2cm}{Lexicographic with warm starts} 
 & $Obj_1, Obj_2$ & 0.0281 & 0.00\%   & 163.46 & 88.95\%  & 86.08 & 35.68\%  & 95.84\%  & 8400.17 & ND \\
 & $Obj_1, Obj_3$ & 0.0281 & 0.00\%   & 183.77 & 100.00\% & 75.32 & 28.32\%  & 103.93\% & 8403.56 & ND \\
 & $Obj_2, Obj_1$ & 0.3220 & 100.00\% & 0.00   & 0.00\%   & 180.2 & 100.00\% & 141.42\% & 0.52    & ND \\
 & $Obj_2, Obj_3$ & 0.3220 & 100.00\% & 0.00   & 0.00\%   & 180.2 & 100.00\% & 141.42\% & 0.56    & ND \\
 & $Obj_3, Obj_1$ & 0.1287 & 34.25\%  & 155.33 & 84.52\%  & 33.88 & 0.00\%   & 91.20\%  & 8400.23 &  ND \\
 & $Obj_3, Obj_2$ & 0.1386 & 37.62\%  & 142.96 & 77.79\%  & 33.88 & 0.00\%   & 86.41\%  & 8400.34 & ND \\ \midrule
\multicolumn{11}{c}{BBCA\_3} \\ 
 \midrule
\multirow{3}{2.1cm}{Single objective optimization} 
 & $Obj_1$ & 0.0402 & 0.00\%   & 193.84 & 118.92\% & 881.35 & 464.42\% & 479.40\% & 4208.84 & D \\
 & $Obj_2$ & 0.6351 & 211.13\% & 0.00   & 0.00\%   & 342.36 & 161.41\% & 265.76\% & 0.58    & D \\
 & $Obj_3$ & 0.6207 & 206.03\% & 172.47 & 105.81\% & 56.09  & 0.48\%   & 231.61\% & 4201.83 & D \\
\midrule
 \multirow{3}{2cm}{Weighted sum} 
 & $Obj_1$ & 0.0402 & 0.00\%  & 140.65 & 86.29\% & 139.88 & 47.58\%  & 98.54\%  & 4212.48 & ND \\
 & $Obj_2$ & 0.2989 & 91.79\% & 0.00   & 0.00\%  & 233.12 & 100.00\% & 135.74\% & 2.69  & ND \\
 & $Obj_3$ & 0.1724 & 46.92\% & 136.37 & 83.66\% & 55.24  & 0.00\%   & 95.92\%  & 4293.23 & ND \\
  \midrule
  \multirow{6}{2cm}{Lexicographic} 
 & $Obj_1, Obj_2$ & \multicolumn{9}{c}{\textit{No feasible solution found}} \\*	
 & $Obj_1, Obj_3$ & \multicolumn{9}{c}{\textit{No feasible solution found}} \\*	
 & $Obj_2, Obj_1$ & 0.2989 & 91.79\% & 0.00 & 0.00\% & 233.12 & 100.00\% & 135.74\% & 0.72  & ND \\
 & $Obj_2, Obj_3$ & 0.3218 & 99.95\% & 0.00 & 0.00\% & 220.48 & 92.89\%  & 136.46\% & 61.82 & ND \\
 & $Obj_3, Obj_1$ & \multicolumn{9}{c}{\textit{No feasible solution found}}  \\*
 & $Obj_3, Obj_2$ & \multicolumn{9}{c}{\textit{No feasible solution found}}  \\	\midrule	
\multirow{6}{2cm}{Lexicographic with warm starts} 
 & $Obj_1, Obj_2$ & 0.0402 & 0.00\%  & 133.40 & 81.84\%  & 150.64 & 53.63\%  & 97.85\%  & 8203.63 & ND \\
 & $Obj_1, Obj_3$ & 0.0402 & 0.00\%  & 160.52 & 98.48\%  & 118.36 & 35.48\%  & 104.68\% & 8405.59 & ND \\
 & $Obj_2, Obj_1$ & 0.2989 & 91.79\% & 0.00   & 0.00\%   & 233.12 & 100.00\% & 135.74\% & 1.12    & ND \\
 & $Obj_2, Obj_3$ & 0.3218 & 99.95\% & 0.00   & 0.00\%   & 220.48 & 92.89\%  & 136.46\% & 58.52   & ND \\
 & $Obj_3, Obj_1$ & 0.1006 & 21.42\% & 163.00 & 100.00\% & 56.09  & 0.48\%   & 102.27\% & 8400.14 & ND \\
 & $Obj_3, Obj_2$ & 0.1897 & 53.04\% & 129.71 & 79.58\%  & 56.09  & 0.48\%   & 95.63\%  & 8400.34 & ND \\
\bottomrule
\end{tabular}
}
\end{sidewaystable}

\FloatBarrier

In the second phase, AUGMECON2 is applied to construct multiobjective solutions in the scenarios with unclassified waste using the approximated ranges of the objectives over the efficient set obtained from the first phase. When using AUGMECON2 for searching a set of feasible solutions in a problem, the upper bound of the number of runs that the solver will performed is determined by the formula $(g+1)^{p-1}$, where $g$ is the number of gridpoints, i.e., the number of intervals in which the objective range is divided that is set by the user, and $p$ is the number of objectives of the problem. This is an upper bound because AUGMECON2 has an acceleration mechanism with early exit from the loops~\citep{mavrotas2013improved}. The number of gridpoints is set at two and, thus, the upper bound of runs is $(2+1)^{3-1} = 9$. 

Table~\ref{tab:MVD_and_BBCA_multiobjective} reports the results for all the proposed scenarios considering unclassified and source classified waste.

\begin{center}
\setlength{\tabcolsep}{2.5pt}
\renewcommand{\arraystretch}{0.85}
\begin{small}
\begin{longtable}{c|rrrrrrrr}
\caption{Multiobjective solutions.\label{tab:MVD_and_BBCA_multiobjective}} \\
\toprule
Solution  & $Obj_1$ & $\Delta Obj_1$ (\%) & $Obj_2$ & $\Delta Obj_1$ (\%) & $Obj_3$ & $\Delta Obj_1$ (\%) & $L_2$ (\%)  & \begin{tabular}{r}Execution\\time (s)\end{tabular} \\ \midrule
\multicolumn{9}{c}{MVD\_1 - unclassified waste} \\* \midrule
1              & 0.3475 & 28.09\% & 0.00   & 0.00\%   & 16600  & 100.00\% & 103.87\% & 0.13    \\*
\textbf{2}              & \textbf{0.1808} & \textbf{2.98\%}  & \textbf{37.76}  & \textbf{33.07\%}  & \textbf{11200}  & \textbf{54.24\%}  & \textbf{63.59\%}  & \textbf{4205.67} \\*
3              & 0.1695 & 1.28\%  & 75.80  & 66.37\%  & 8400   & 30.51\%  & 73.06\%  & 4203.55 \\*
4              & 0.1638 & 0.43\%  & 112.78 & 98.76\%  & 7500   & 22.88\%  & 101.37\% & 4200.35 \\ \midrule
\multicolumn{9}{c}{MVD\_2 - unclassified waste} \\* \midrule
1              & 0.1016 & 0.00\%  & 110.78 & 97.01\%  & 4500   & 13.83\%  & 97.99\%  & 0.13    \\*
2              & 0.1016 & 0.00\%  & 74.11  & 64.89\%  & 5500   & 24.47\%  & 69.35\%  & 4203.56 \\*
\textbf{3}              & \textbf{0.1120} & \textbf{2.06\%}  & \textbf{37.64}  & \textbf{32.96\%}  & \textbf{8900}   & \textbf{60.64\%}  & \textbf{69.05\%}  & \textbf{4233.70} \\*
4              & 0.3333 & 45.88\% & 0.00   & 0.00\%   & 12600  & 100.00\% & 110.02\% & 4206.60 \\ \midrule
\pagebreak 
\midrule
\multicolumn{9}{c}{MVD\_3 - unclassified waste} \\* \midrule
1              & 0.0452 & 0.00\%  & 129.76 & 113.62\% & 2400   & 12.12\%  & 114.27\% & 4205.60 \\*
2              & 0.0571 & 3.52\%  & 131.30 & 114.97\% & 1600   & 0.00\%   & 115.03\% & 4208.30 \\*
3              & 0.0452 & 0.00\%  & 92.21  & 80.75\%  & 3400   & 27.27\%  & 85.23\%  & 4203.50 \\*
4              & 0.0667 & 6.34\%  & 45.69  & 40.00\%  & 4900   & 50.00\%  & 64.35\%  & 4204.10 \\*
\textbf{5}              & \textbf{0.0857} & \textbf{11.97\%} & \textbf{46.57}  & \textbf{40.78\%}  & \textbf{3800}   & \textbf{33.33\%}  & \textbf{54.01\%}  & \textbf{4206.90} \\*
6              & 0.3333 & 85.21\% & 0.00   & 0.00\%   & 8200   & 100.00\% & 131.38\% & 0.14    \\ \midrule
\multicolumn{9}{c}{BBCA\_1 - unclassified waste} \\* \midrule
1              & 0.0909 & 1.02\%  & 152.71 & 97.72\%  & 146.24 & 28.06\%  & 101.68\% & 4202.20 \\*
2              & 0.0966 & 2.04\%  & 102.61 & 65.67\%  & 226.63 & 64.48\% & 92.06\% & 4200.40 \\*
\textbf{3}              & \textbf{0.1174} & \textbf{5.78\%}  & \textbf{103.88} & \textbf{66.48\%}  & \textbf{156.24} & \textbf{32.59\%}  & \textbf{74.26\%} & \textbf{4200.00} \\*
4              & 0.3333 & 44.56\% & 0.00   & 0.00\%   & 305.02 & 100.00\% & 109.48\% & 1.51    \\*
5              & 0.4451 & 64.63\% & 0.00   & 0.00\%   & 227.63 & 64.94\% & 91.96\% & 1601.27 \\ \midrule
\multicolumn{9}{c}{BBCA\_2 - unclassified waste} \\* \midrule
1              & 0.0297 & 0.56\%  & 153.24 & 83.39\%  & 103.20 & 47.38\%  & 95.91\%  & 4206.00 \\*
2              & 0.0495 & 7.30\%  & 101.80 & 55.40\%  & 99.20  & 44.64\%  & 71.52\%  & 4207.40 \\*
\textbf{3}              & \textbf{0.1089} & \textbf{27.51\%} & \textbf{49.95}  & \textbf{27.18\%}  & \textbf{116.20} & \textbf{56.26\%}  & \textbf{68.27\%}  & \textbf{4205.00} \\ \midrule
\multicolumn{9}{c}{BBCA\_3 - unclassified waste} \\* \midrule
1              & 0.0402 & 0.00\%  & 160.10 & 98.22\%  & 129.12 & 41.53\%  & 106.64\% & 4206.30 \\*
\textbf{2}              & \textbf{0.0690} & \textbf{10.20\%} & \textbf{106.97} & \textbf{65.62\%}  & \textbf{172.12} & \textbf{65.71\%}  & \textbf{93.42\%}  & \textbf{4203.60} \\*
3              & 0.2989 & 91.79\% & 0.00   & 0.00\%   & 233.12 & 100.00\% & 135.74\% & 3.92  \\ \midrule
\multicolumn{9}{c}{MVD\_1 - classified waste} \\* \midrule
1              & 0.1879 & 0.80\%  & 127.22 & 95.69\% & 9900   & 36.92\%  & 102.57\% & 4200.12 \\*
2              & 0.1864 & 0.40\%  & 84.93  & 63.88\% & 11300  & 47.69\%  & 79.72\%  & 4202.34 \\*
\textbf{3}              & \textbf{0.1822} & \textbf{-0.80\%} & \textbf{44.29}  & \textbf{33.31\%} & \textbf{10500}  & \textbf{41.54\%}  & \textbf{53.25\%}  & \textbf{4203.12} \\*
4              & 0.3475 & 46.18\% & 0.00   & 0.00\%  & 18100  & 100.00\% & 110.15\% & 0.16     \\ \midrule
\multicolumn{9}{c}{MVD\_2 - classified waste} \\* \midrule
1              & 0.1094 & -3.33\% & 123.36 & 81.28\% & 8500   & 38.81\%  & 90.13\%  & 4225.40 \\*
2              & 0.1120 & -2.50\% & 96.86  & 63.82\% & 8000   & 35.07\%  & 72.87\%  & 4225.40 \\*
\textbf{3}              & \textbf{0.1133} & \textbf{-2.08\%} & \textbf{50.09}  & \textbf{33.01\%} & \textbf{9400}   & \textbf{45.52\%}  & \textbf{56.27\%}  & \textbf{4225.40} \\*
4              & 0.3346 & 68.75\% & 76.51  & 50.41\% & 16700  & 100.00\% & 131.41\% & 4225.40 \\ \midrule
\pagebreak 
\midrule
\multicolumn{9}{c}{MVD\_3 - classified waste} \\* \midrule
1              & 0.0500 & 0.41\%  & 121.00 & 86.20\% & 5500   & 30.71\%  & 91.51\%  & 4211.70 \\*
2             & 0.0524& 1.24\% & 92.31 & 65.76\%& 5300  & 29.13\% & 71.94\% & 4219.70\\*
\textbf{3}              & \textbf{0.0667} & \textbf{6.20\%}  & \textbf{46.64}  & \textbf{33.23\%} & \textbf{5500}   & \textbf{30.71\%}  & \textbf{45.67\%}  & \textbf{4285.90} \\*
4              & 0.3333 & 98.76\% & 0.00   & 0.00\%  & 14300  & 100.00\% & 140.55\% & 0.16    \\ \midrule
\multicolumn{9}{c}{BBCA\_1 - classified waste} \\* \midrule
\textbf{1}              & \textbf{0.2528} & \textbf{56.76\%} & \textbf{54.48}  & \textbf{38.81\%} & \textbf{397.89} & \textbf{96.16\%}  & \textbf{118.21\%} & \textbf{4200.00} \\*
2              & 0.3333 & 85.47\% & 0.00   & 0.00\%  & 409.89 & 100.00\% & 131.55\% & 0.36    \\ \midrule
\multicolumn{9}{c}{BBCA\_2 - classified waste} \\* \midrule
1              & 0.0429 & 3.37\%  & 167.08 & 97.68\% & 124.00 & 21.65\%  & 100.11\% & 4230.50 \\*
2              & 0.0462 & 4.33\%  & 113.39 & 66.29\% & 123.00 & 21.38\%  & 69.79\%  & 4235.70 \\*
\textbf{3}              & \textbf{0.0990} & \textbf{19.74\%} & \textbf{56.73}  & \textbf{33.17\%} & \textbf{158.00} & \textbf{30.79\%}  & \textbf{49.37\%}  & \textbf{19.90}   \\ \midrule
\multicolumn{9}{c}{BBCA\_3 - classified waste} \\* \midrule
1              & 0.0488 & 5.09\%  & 140.00 & 81.85\% & 118.00 & 13.41\%  & 83.09\%  & 4221.00 \\*
2              & 0.0571 & 7.53\%  & 100.27 & 58.62\% & 108.00 & 10.73\%  & 60.07\%  & 4283.80 \\*
\textbf{3}              & \textbf{0.1048} & \textbf{21.42\%} & \textbf{49.54}  & \textbf{28.96\%} & \textbf{135.00} & \textbf{17.95\%}  & \textbf{40.25\%}  & \textbf{4240.20} \\*
4              & 0.2952 & 77.00\% & 0.00   & 0.00\%  & 265.00 & 52.69\%  & 93.30\%  & 3.31   \\
\bottomrule
\end{longtable}
\end{small}
\end{center}

\FloatBarrier

\subsection{Experimental evaluation: PageRank heuristic}
\label{subsec:resultsHeuristic}

Table~\ref{tab:pagerank} reports the results of the proposed PageRank heuristic algorithms for the scenarios of Bah\'ia Blanca and Montevideo. In all the scenarios the PageRank algorithms provided solutions that collected all the waste even though this was not \green{imposed} in the formulation. The first column indicates the variant of PageRank heuristic used, then the value of the optimized objectives of distance and installment cost are reported, and finally the execution times (in seconds) are presented.

\begin{table}[!h]
\caption{Solutions obtained by using Pagerank methods in Bah\'{i}a Blanca.\label{tab:pagerank}}
\centering
\setlength{\tabcolsep}{16pt}
\renewcommand{\arraystretch}{0.85}
{\small 
\begin{tabular}{l rrrr}
\toprule
\multicolumn{1}{c}{\textit{Method}} & $Obj_2$ & $Obj_3$ & Execution time (s) \\ \midrule
\multicolumn{4}{c}{MVD\_1} \\ \midrule
PageRank-Cost & 100.30 & 5400 & 5.74 \\
PageRank-Dist & 0.00 & 12700 & 6.32 \\
PageRank-Vol & 174.57 & 7900 & 5.77 \\
\midrule
\multicolumn{4}{c}{MVD\_2} \\ \midrule
PageRank-Cost & 98.44 & 3400 & 7.46 \\
PageRank-Dist & 0.00 & 13700 & 8.80 \\
PageRank-Vol  & 166.52 & 4000 & 6.76 \\
\midrule
\multicolumn{4}{c}{MVD\_3} \\ \midrule
PageRank-Cost & 123.39 & 1700 & 8.58 \\
PageRank-Dist & 0.00 & 14100 & 10.48 \\
PageRank-Vol & 176.27 & 2800 & 8.01 \\
\midrule
\multicolumn{4}{c}{BBCA\_1} \\ \midrule
PageRank-Cost & 155.85 & 122.96 & 11.23\\
PageRank-Dist &  0.00 & 380.53 & 12.34\\
PageRank-Vol &  209.14 & 196.42 & 11.23\\
\midrule
\multicolumn{4}{c}{BBCA\_2} \\ \midrule
PageRank-Cost &   191.01 & 55.12 & 17.83\\
PageRank-Dist & 0.00 & 428.24 & 18.81\\
PageRank-Vol & 222.59 & 98.51 & 16.28\\
\midrule
\multicolumn{4}{c}{BBCA\_3} \\ \midrule
PageRank-Cost &  162.51 & 82.68 & 21.27\\
PageRank-Dist & 0.00 & 492.89 & 25.12\\
PageRank-Vol  & 204.52 & 148.52 & 20.69\\ 
\bottomrule
\end{tabular}
}
\end{table}

\subsection{Analysis of results}
\label{subsec:analysis_results}

From the first phase of the exact approach, some results are important to highlight. The problems that minimize the average distance ($Obj_2$) (or that considered minimizing $Obj_2$ in the first stage) are easier to solve and optimal solutions are found, in most cases in less than 5 seconds. Straightforward lexicographic optimization can only find feasible solutions when $Obj_2$ is optimized on the first stage. However, when warm starts are used, lexicographic optimization is always able to find feasible solutions within the considered time limit in all executions. Moreover, these solutions are generally non-dominated by single objective optimization and the weighted sum. In some particular scenarios the weighted sum was able to find non-dominated solutions too, e.g., three solutions in scenario {BBCA\_3}. These exceptions are valuable, considering that weighted sum is less time consuming than lexicographic approaches. The second phase computed some compromise solutions that can be useful for decision makers to not depend on the extreme solutions computed in the first phase. However, the number of feasible solutions obtained in some scenarios is quite \green{small considering} that the upper bound for the number of runs to perform is 9. This results is due 
to \green{the} NP-hard nature of the problem, which causes that no feasible \green{solutions} are found within the time limit in several trials. Another important aspect is that AUGMECON2 was able to find solutions that improved the ideal value of $Obj_1$ in scenarios MVD\_1 and MVD\_2 with classified waste.

PageRank-Cost was the best heuristic, consistently outperforming PageRank-Dist and PageRank-Vol in all problem scenarios, even for QoS-related metrics considered in the study. Although PageRank methods are not forced to collect all the waste, they provided solutions that collected all the available waste. Solutions obtained with the exact method outperformed PageRank-Cost in terms of installment cost in all scenarios but MVD\_2. They also provided solutions that dominated the solution of PageRank-Dist with the same value of $Obj_2$ but better values of cost . The installment cost of PageRank-Cost solutions were between 6.25\% and 12.50\% higher for  Montevideo scenarios and from 44.91\% and 62.69\% higher for Bah\'ia Blanca scenarios.
Nevertheless, the heuristic solutions where obtained, on average, with remarkably shorter computing times.

\begin{figure}[ht]
  	\includegraphics[width=1\textwidth]{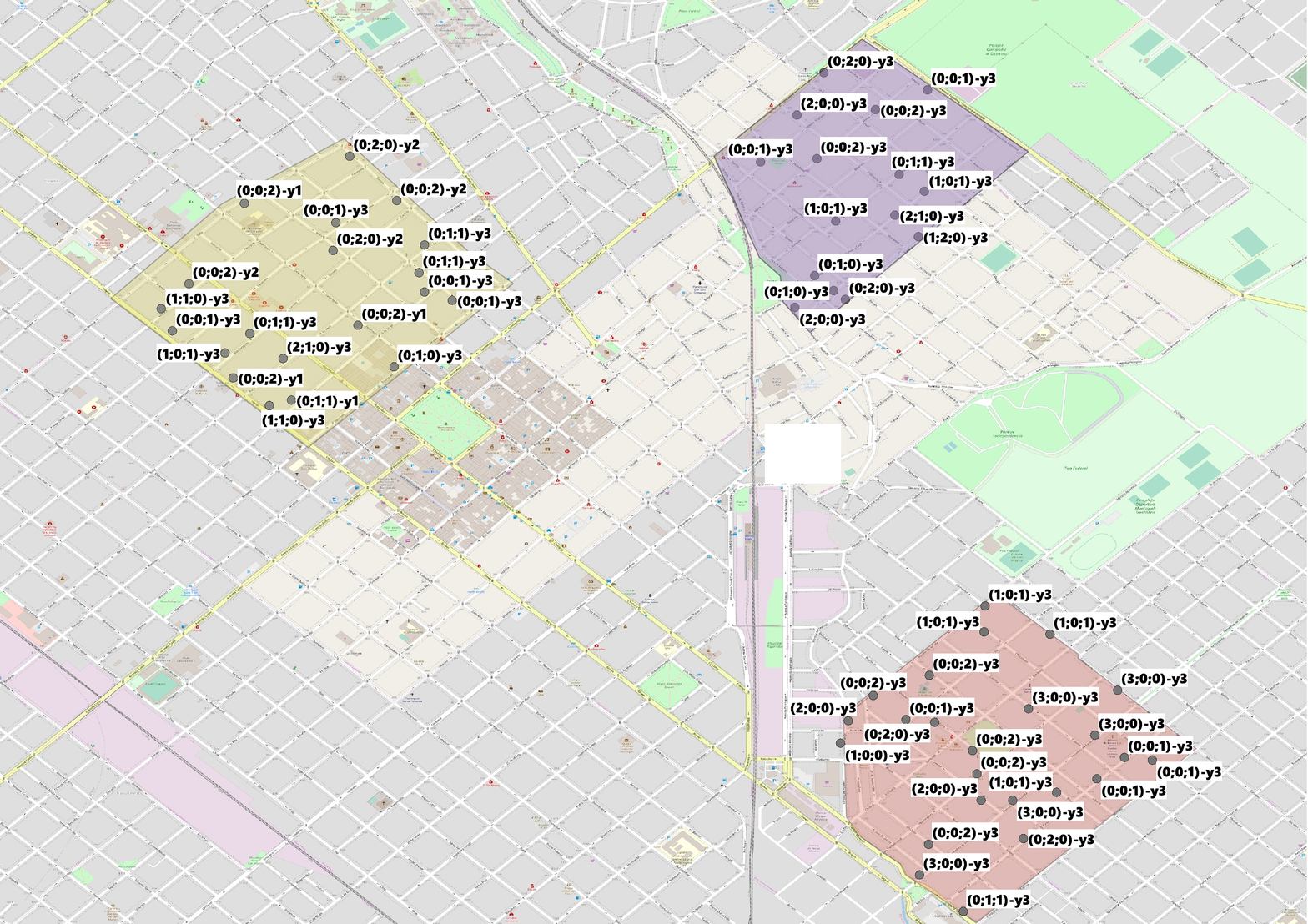}
    \vspace{-0.4cm}
  \caption{\blue{Used GAPs in candidate solutions for the scenarios of BBCA\_1 (yellow), BBCA\_2 (blue) and BBCA\_3 (red) with unclassified waste. Each GAP is labeled with a vector representing the bin configuration ($j_1$, $j_2$, $j_3$) and the assigned collection frequency. Source: background map from~\cite{OpenStreetMap}.} \label{fig:Result_BBCA}}
\end{figure}

\begin{figure}[ht]
  \begin{center}
  \begin{minipage}{.475\linewidth}
  	\includegraphics[width=1\textwidth]{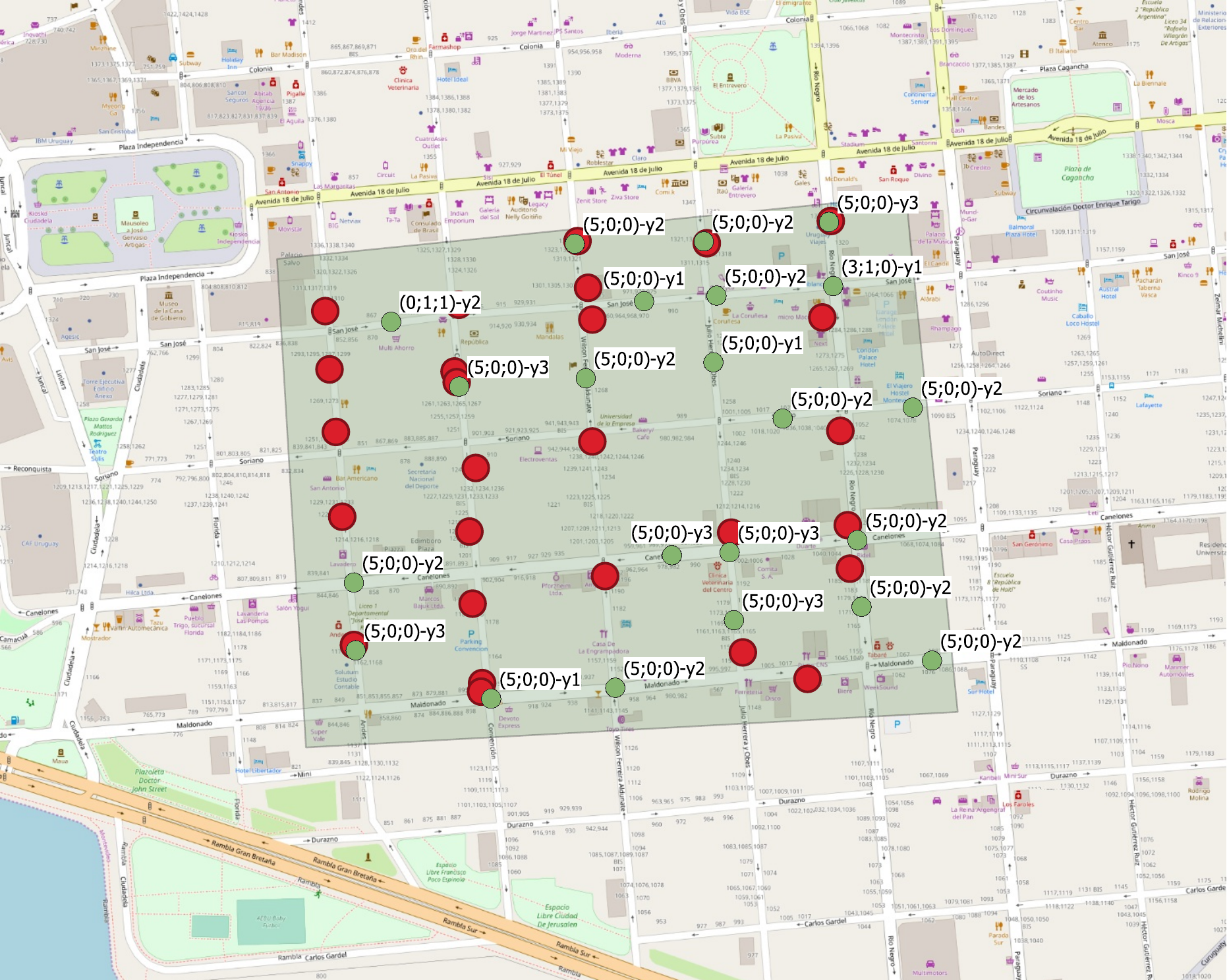}
  \end{minipage}
  \begin{minipage}{.475\linewidth}
    \includegraphics[width=1\textwidth]{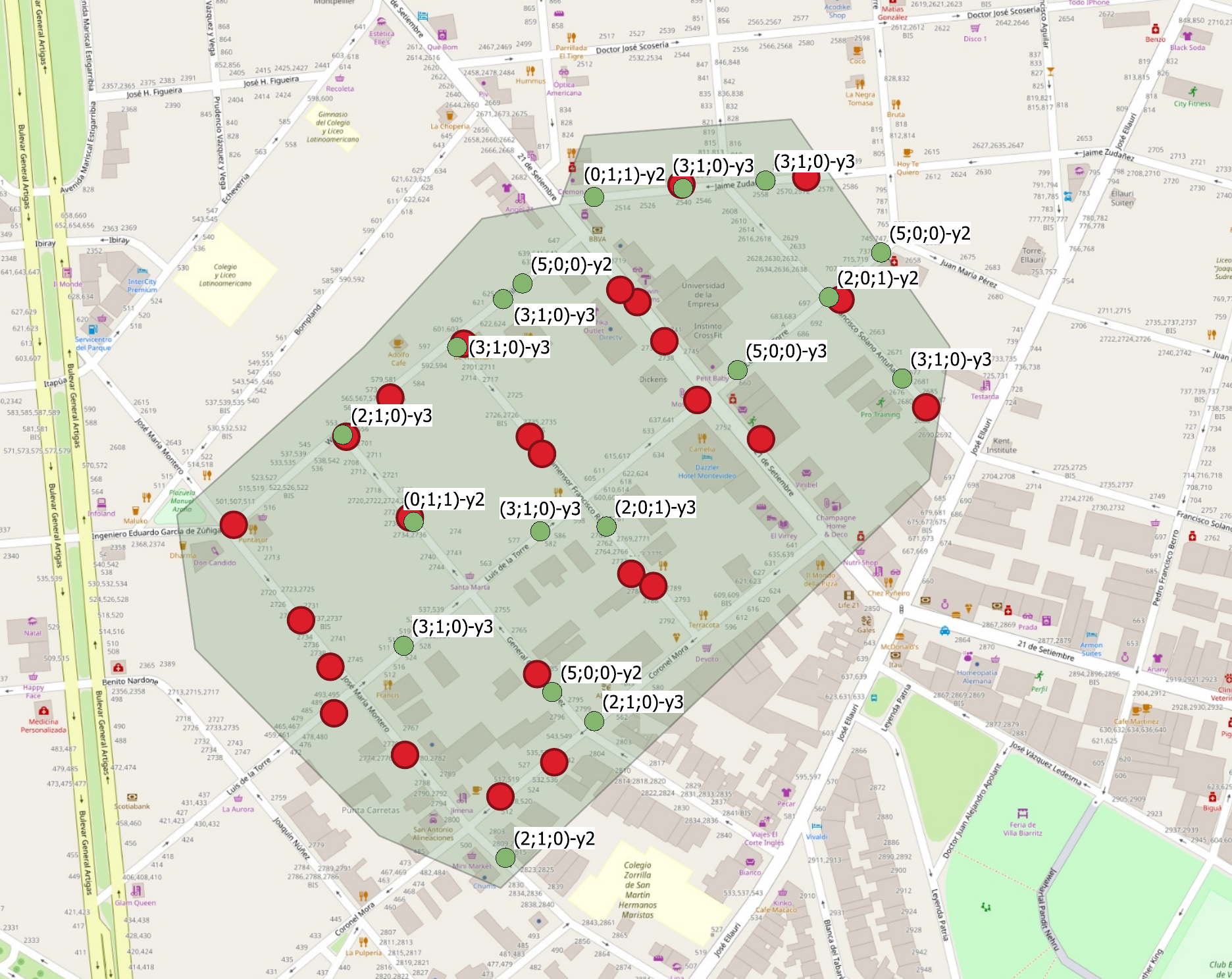}
  \end{minipage}
  \end{center}
  \vspace{-0.4cm}
  \caption{\blue{Used GAPs (green circles) in candidate solutions for the scenarios of MVD\_1 (left) and MVD\_2 (right) with classified waste. Each GAP is labeled with a vector representing the bin configuration ($j_1$, $j_2$, $j_3$) and the assigned collection frequency. Red circles are current real location of GAPs. Source: background map from~\cite{OpenStreetMap}.}\label{fig:Result_MVD_1_2}}
\end{figure}

\blue{The aim of multiobjective optimization is to present a representative set of solutions that aims to reflect the trade-offs among the different optimized criteria. Then, if a candidate solution is to be selected, a common practice is to pick the solution that is nearest to the ideal multiobjective solution~\citep{deb2001multi}. In this case, the ones that have the smaller $L^2$ which are highlighted in Table~\ref{tab:MVD_and_BBCA_multiobjective} for every scenario. In Fig.~\ref{fig:Result_BBCA}, the candidate solutions for the three scenarios of Bah\'{i}a Blanca with unclassified waste are depicted. Then, in Fig.~\ref{fig:Result_MVD_1_2} candidate solutions for classified waste scenarios of MVD\_1 and MVD\_2 are shown. In the particular case of Bah\'{i}a Blanca, the approaches allowed to obtain some feasible solutions for the GAP location problem in the city \green{while} considering two costs measures such as the number of installed bins and minimizing the posterior routing cost. This last criterion is of particular importance for Argentinian cities since they experience one of the highest logistic \green{costs} in the region~\citep{broz2018argentinian} and, thus, solutions with particularly small $\Delta Obj_1$ may also be of interest for the City Hall of Bah\'ia Blanca to implement the community bins system}.

\green{In the case of Montevideo, the candidate solutions are useful for comparison with the current GAPs distribution. Therefore, they were compared with simulations performed with the current location of GAPs in Montevideo. Central neighborhoods of Montevideo, such as MVD\_1, have community bins for both recyclable and mixed waste. However, in other neighborhoods such as MVD\_2 and MVD\_3, recyclable waste is deposited in rather sparse large bins located in crowded places~\green{\citep{manualresiduos}}. To perform a fair comparison, in these cases bins of recyclable waste were supposed to be located next to mixed waste ones. The improvements of the candidate solutions over the current locations are reported in Table~\ref{tab:Real_MVD}. Overall, the candidate solutions substantially improved results in terms of frequency collection (up to 51.1\% of improvement in instance MVD$\_1$) and average distance to the users (up to 49.2\% of improvement in instance MVD$\_3$). On the one hand, these results indicate that the proposed model is able to accurately compute solutions with a significantly better QoS, with a direct implication for citizens. On the other hand, for the cost objective improvements were obtained for two out of the three studied scenarios. In MVD$\_1$ the cost of the candidate solution is larger than the one from the real current locations of GAPs. This is a busy area with several touristic places and public buildings and, thus, it is important to avoid the negative impact of overflowed bins. The City Hall empties the bins on a daily basis and, therefore, the installed capacity (related to number and types of bins) required for accumulating waste in the GAPs is relatively small (only receives the amount generated within a day). On the other hand, in the candidate solution, where GAPs are generally emptied every two or three days (see Fig.~\ref{fig:Result_MVD_1_2}),  larger investment cost in GAPs' capacity is required (since they need to store the waste generated during two or three days). This evinces the aforementioned compromising solution of the GAP location costs and the posterior operational routing costs. \magenta{In the other scenarios, i.e., MVD\_2 and MVD\_3, waste is collected by the City Hall with a frequency that has a maximum number of days between two consecutive visits of three days ($a_{y_{3}}=3$)}.}

\begin{table}
\setlength{\tabcolsep}{12pt}
\centering
\begin{small}
\caption{\green{Comparison between candidate multiobjective solutions and current real location of GAPs in Montevideo.}\label{tab:Real_MVD}} 
\begin{tabular}{c|rrr}
\toprule
\multirow{2}{*}{Instance} & \multicolumn{3}{c}{Objective}\\
\cline{2-4}
& $Obj_1$ & $Obj_2$ & $Obj_3$  \\ \midrule
MVD$\_1$         & -51.14\% & -37.76\% & 87.5\% \\*
MVD$\_2$         & -44.93\% & -23.92\%  & -12.15\% \\*
MVD$\_3$         & -27.25\% & -49.21\%  & -1.78\% \\
\bottomrule
\end{tabular}
\end{small}
\end{table}

\FloatBarrier

\section{Conclusions and future work}
\label{sec:conclusion}

This article studied the problem of locating community waste bins in urban areas. This is a relevant problem for decision makers of cities that already implement a community bins system, as Montevideo, Uruguay, or are willing to implement it, as Bah\'{i}a Blanca, Argentina. Real-world instances of these two cities were used in the experimentation.

Exact and heuristic approaches were developed to solve the problem. The exact model aims at optimizing the investment cost, the accessibility to the system, and a proxy indicator of the routing cost (the collection frequency required to remove waste from the bins). This last objective has not been usually addressed in the literature. \blue{However, it is a relevant criteria for cities that experience remarkably large transport costs as occurs for some of the analyzed scenarios}. The exact algorithm is based \green{on} AUGMECON, which uses as input the ideal and nadir values of each objective. Four methods for obtaining the ranges were compared: single objective optimization, unbalanced weighted sum, straightforward lexicographic optimization, and a novel variation that uses a warm start to enhance lexicographic optimization. Results indicate that including warm starts remarkably increases the convergence of the lexicographic approach. In turn, weighted sum also obtained some relevant results for particular scenarios. \green{Moreover, selected multiobjective solutions were compared with simulations performed with the current GAP distribution in Montevideo, obtaining improvements up to 51.14\% in terms of cost, up to 37.76\% in terms of accessibility to the system, and up to 12.15\% in terms of collection frequency.}
A family of PageRank heuristics were also developed to construct feasible solutions while optimizing the same first and second objective of the exact approach and a third objective to maximize the collected waste. This is a valuable approach in developing countries where formal MSW system usually competes with informal collection. Results show that PageRank can quickly generate some feasible solutions for the GAP location problem.

Having different efficient and accurate methods for optimization is important for providing to decision-makers a set of candidate solutions to analyze, taking into account the different trade-offs between the desired criteria. Therefore, the main lines for future work are related to further study efficient exact methods and heuristics for solving the GAP location problem, especially considering larger scenarios. For this purpose, the use of metaheuristics is a promising line of work to improve the applicability of the proposed model. Regarding exact methods, extending the capacity of computing a representative set of feasible multiobjective solutions within reasonable computing times is important for validating the heuristics in reduced scenarios. Additionally, the presented models can be extended to include some other relevant aspects of the reality, such as considering \green{a} stochastic waste generation rate. \magenta{Finally, another research line is to consider a larger variety of scenarios from different cities, in which administrative, commercial and residential divisions are clearer, in order to continue analyzing the performance of the proposed methods.}.

\section*{Acknowledgements}
We would like to thank the anonymous reviewers for their insightful comments on the paper that led us to a substantial improvement of this work. 
J. Toutouh research was partially funded by European Union’s Horizon 2020 research and innovation program under the Marie Skłodowska-Curie grant agreement No 799078 and the Systems that Learn Initiative at MIT CSAIL.

\bibliographystyle{abbrvnat}
\bibliography{mybibfile}

\appendix

\setcounter{table}{0}
\renewcommand{\thetable}{A\arabic{table}}

\section{\green{First stage results for classified waste}}
\label{appendix:first_stage}

\green{Tables~\ref{tab:obj_range_MVDI_and_MVDII_clas}--\ref{tab:obj_range_BBCAII_and_BBCAIII_clas} report the results of the experimental evaluation for approximating the efficient range the objectives in the classified waste scenarios of Montevideo and Bah\'ia Blanca.}

\begin{sidewaystable}
\centering
\setlength{\tabcolsep}{4pt}
\renewcommand{\arraystretch}{0.7}
\caption{\green{Objectives range approximation for scenarios of MVD\_1 and MVD\_2 with classified waste.}\label{tab:obj_range_MVDI_and_MVDII_clas}}
{\scriptsize 
\begin{tabular}{cc rrr rrr rr c}
\toprule
Method & \begin{tabular}{c} Optimized\\objective \end{tabular} & $Obj_1$ & $\Delta Obj_1$ (\%) & $Obj_2$ & $\Delta Obj_1$ (\%) & $Obj_3$ & $\Delta Obj_1$ (\%) & $L_2$ (\%)  & \begin{tabular}{r}Execution\\time (s)\end{tabular}  & Dominance \\ \midrule
\multicolumn{11}{c}{MVD\_1} \\ \midrule
\multirow{3}{2.1cm}{Single objective optimization} 
 & $Obj_1$ & 0.1850 & 0.00\%   & 199.71 & 150.22\% & 29500 & 187.69\% & 240.40\% & 4206 & D \\*
 & $Obj_2$ & 0.9732 & 224.10\% & 0.00   & 0.00\%   & 29500 & 187.69\% & 292.31\% & 2    & D \\*
 & $Obj_3$ & 0.3347 & 42.57\%  & 211.41 & 159.01\% & 5100  & 0.00\%   & 164.61\% & 4203 & D \\ \midrule
 \multirow{3}{2cm}{Weighted sum} 
 & $Obj_1$ & 0.1921 & 2.01\%  & 162.59 & 122.30\% & 9500  & 33.85\% & 126.91\% & 4205 & D \textbackslash{}\textbackslash{}* \\
 & $Obj_2$ & 0.3898 & 58.23\% & 0.00   & 0.00\%   & 15300 & 78.46\% & 97.71\%  & 3    & ND                                  \\
 & $Obj_3$ & 0.2203 & 10.04\% & 123.46 & 92.86\%  & 5100  & 0.00\%  & 93.40\%  & 4200 & ND  \\
  \midrule
  \multirow{6}{2cm}{Lexicographic} 
 & $Obj_1, Obj_2$ & \multicolumn{9}{c}{\textit{No feasible solution found}} \\*	
 & $Obj_1, Obj_3$ & \multicolumn{9}{c}{\textit{No feasible solution found}} \\*	
 & $Obj_2, Obj_1$ & 0.3475 & 46.18\%  & 0.00 & 0.00\% & 18100 & 100.00\% & 110.15\% & 0 & ND \\
 & $Obj_2, Obj_3$ & 0.5367 & 100.00\% & 0.00 & 0.00\% & 12400 & 56.15\%  & 114.69\% & 0 & ND \\*
 & $Obj_3, Obj_1$ & \multicolumn{9}{c}{\textit{No feasible solution found}}  \\*
 & $Obj_3, Obj_2$ & \multicolumn{9}{c}{\textit{No feasible solution found}}  \\	\midrule	
\multirow{6}{2cm}{Lexicographic with warm starts} 
 & $Obj_1, Obj_2$ & 0.1879 & 0.80\%   & 37.53  & 28.23\%  & 13300 & 63.08\%  & 69.11\%  & 8405 & ND \\
 & $Obj_1, Obj_3$ & 0.1864 & 0.40\%   & 132.95 & 100.00\% & 5300  & 1.54\%   & 100.01\% & 4200 & ND \\
 & $Obj_2, Obj_1$ & 0.3475 & 46.18\%  & 0.00   & 0.00\%   & 18100 & 100.00\% & 110.15\% & 1    & ND \\
 & $Obj_2, Obj_3$ & 0.5367 & 100.00\% & 0.00   & 0.00\%   & 12400 & 56.15\%  & 114.69\% & 1    & ND \\
 & $Obj_3, Obj_1$ & 0.2542 & 19.68\%  & 103.69 & 77.99\%  & 5100  & 0.00\%   & 80.44\%  & 8405 & ND \\
 & $Obj_3, Obj_2$ & 0.2881 & 29.32\%  & 77.27  & 58.12\%  & 5100  & 0.00\%   & 65.10\%  & 8403 & ND \\ \midrule
\multicolumn{11}{c}{MVD\_2} \\ 
\midrule
\multirow{3}{2.1cm}{Single objective optimization} 
 & $Obj_1$ & 0.1198 & 0.00\%   & 181.25 & 119.43\% & 32000 & 214.18\% & 245.23\% & 4207 & D  \\
 & $Obj_2$ & 0.6120 & 157.50\% & 0.00   & 0.00\%   & 32000 & 214.18\% & 265.86\% & 3    & D  \\
 & $Obj_3$ & 0.2839 & 52.50\%  & 192.00 & 126.51\% & 3300  & 0.00\%   & 136.97\% & 4203 & ND \\
\midrule
 \multirow{3}{2cm}{Weighted sum} 
 & $Obj_1$ & 0.1263 & 2.08\%  & 145.25 & 95.71\% & 7800  & 33.58\% & 101.45\% & 4209 & ND \\
 & $Obj_2$ & 0.3646 & 78.33\% & 0.00   & 0.00\%  & 14400 & 82.84\% & 114.01\% & 3    & ND \\
 & $Obj_3$ & 0.1563 & 11.67\% & 105.70 & 69.65\% & 3400  & 0.75\%  & 70.62\%  & 4200 & ND \\
  \midrule
  \multirow{6}{2cm}{Lexicographic} 
 & $Obj_1, Obj_2$ & \multicolumn{9}{c}{\textit{No feasible solution found}} \\*	
 & $Obj_1, Obj_3$ & \multicolumn{9}{c}{\textit{No feasible solution found}} \\*	
 & $Obj_2, Obj_1$ & 0.3346 & 68.75\%  & 0.00 & 0.00\% & 16700 & 100.00\% & 121.35\% & 0 & ND \\
 & $Obj_2, Obj_3$ & 0.4323 & 100.00\% & 0.00 & 0.00\% & 12900 & 71.64\%  & 123.01\% & 0 & ND \\
 & $Obj_3, Obj_1$ & \multicolumn{9}{c}{\textit{No feasible solution found}}  \\*
 & $Obj_3, Obj_2$ & \multicolumn{9}{c}{\textit{No feasible solution found}}  \\	\midrule	
\multirow{6}{2cm}{Lexicographic with warm starts} 
 & $Obj_1, Obj_2$ & 0.1198 & 0.00\%   & 33.02   & 21.75\%  & 10300 & 52.24\%  & 56.59\%  & 8403 & ND \\
 & $Obj_1, Obj_3$ & 0.1198 & 0.00\%   & 151.766 & 100.00\% & 3600  & 2.24\%   & 100.03\% & 8402 & ND \\
 & $Obj_2, Obj_1$ & 0.3346 & 68.75\%  & 0.00    & 0.00\%   & 16700 & 100.00\% & 121.35\% & 0    & ND \\
 & $Obj_2, Obj_3$ & 0.4323 & 100.00\% & 0.00    & 0.00\%   & 12900 & 71.64\%  & 123.01\% & 1    & ND \\
 & $Obj_3, Obj_1$ & 0.1719 & 16.67\%  & 145.72  & 96.02\%  & 3300  & 0.00\%   & 97.45\%  & 8401 & D  \\
 & $Obj_3, Obj_2$ & 0.1719 & 16.67\%  & 134.41  & 88.56\%  & 3300  & 0.00\%   & 90.12\%  & 8415 & ND \\
\bottomrule
\end{tabular}
}
\end{sidewaystable}

\begin{sidewaystable}
\centering
\setlength{\tabcolsep}{4pt}
\renewcommand{\arraystretch}{0.7}
{\scriptsize 
\caption{\green{Objectives range approximation for scenarios of MVD\_3 and BBCA\_1 with classified waste.}\label{tab:obj_range_MVDIII_and_BBCAI_clas}}
\begin{tabular}{cc| rrrrrrrrc}
\toprule
Method & \begin{tabular}{c} Optimized\\objective \end{tabular} & $Obj_1$ & $\Delta Obj_1$ (\%) & $Obj_2$ & $\Delta Obj_1$ (\%) & $Obj_3$ & $\Delta Obj_1$ (\%) & $L_2$ (\%)  & \begin{tabular}{r}Execution\\time (s)\end{tabular}  & Dominance \\ \midrule
\multicolumn{11}{c}{MVD\_3} \\ \midrule
\multirow{3}{2.1cm}{Single objective optimization} 
 & $Obj_1$ & 0.0488 & 0.00\%   & 192.16 & 136.90\% & 35000 & 262.99\% & 296.49\% & 4200 & D \\
 & $Obj_2$ & 0.6310 & 202.07\% & 0.00   & 0.00\%   & 35000 & 262.99\% & 331.66\% & 1    & D \\
 & $Obj_3$ & 0.2179 & 58.68\%  & 193.10 & 137.57\% & 1800  & 1.57\%   & 149.57\% & 4204 & D \\ \midrule
 \multirow{3}{2cm}{Weighted sum} 
 & $Obj_1$ & 0.0488 & 0.00\%  & 115.63 & 82.38\% & 4500  & 22.83\% & 85.48\%  & 4214 & ND \\
 & $Obj_2$ & 0.3345 & 99.17\% & 0.00   & 0.00\%  & 14200 & 99.21\% & 140.28\% & 4    & ND \\
 & $Obj_3$ & 0.0738 & 8.68\%  & 118.19 & 84.20\% & 1600  & 0.00\%  & 84.64\%  & 4216 & ND \\
  \midrule
  \multirow{6}{2cm}{Lexicographic} 
 & $Obj_1, Obj_2$ & \multicolumn{9}{c}{\textit{No feasible solution found}} \\*	
 & $Obj_1, Obj_3$ & \multicolumn{9}{c}{\textit{No feasible solution found}} \\*	
 & $Obj_2, Obj_1$ & 0.3333 & 98.76\%  & 0.00 & 0.00\% & 14300 & 100.00\% & 140.55\% & 0 & ND \\
 & $Obj_2, Obj_3$ & 0.3369 & 100.00\% & 0.00 & 0.00\% & 14000 & 97.64\%  & 139.76\% & 0 & ND \\*
 & $Obj_3, Obj_1$ & \multicolumn{9}{c}{\textit{No feasible solution found}}  \\*
 & $Obj_3, Obj_2$ & \multicolumn{9}{c}{\textit{No feasible solution found}}  \\	\midrule	
\multirow{6}{2cm}{Lexicographic with warm starts} 
 & $Obj_1, Obj_2$ & 0.0488 & 0.00\%   & 67.77  & 48.28\%  & 5000  & 26.77\%  & 55.21\%  & 8405 & ND \\
 & $Obj_1, Obj_3$ & 0.0488 & 0.00\%   & 160.37 & 114.25\% & 3500  & 14.96\%  & 115.22\% & 8404 & D  \\
 & $Obj_2, Obj_1$ & 0.3333 & 98.76\%  & 0.00   & 0.00\%   & 14300 & 100.00\% & 140.55\% & 1    & ND \\
 & $Obj_2, Obj_3$ & 0.3369 & 100.00\% & 0.00   & 0.00\%   & 14000 & 97.64\%  & 139.76\% & 1    & ND \\
 & $Obj_3, Obj_1$ & 0.0476 & -0.41\%  & 140.37 & 100.00\% & 1600  & 0.00\%   & 100.00\% & 8401 & ND \\
 & $Obj_3, Obj_2$ & 0.1143 & 22.73\%  & 121.90 & 86.84\%  & 1800  & 1.57\%   & 89.78\%  & 8411 & D \\ \midrule
\multicolumn{11}{c}{BBCA\_1} \\ 
 \midrule
\multirow{3}{2.1cm}{Single objective optimization} 
 & $Obj_1$ & 0.0938 & 0.00\%   & 175.28 & 124.87\% & 509.80 & 131.97\% & 181.68\% & 4207 & D \\
 & $Obj_2$ & 0.6184 & 187.16\% & 0.00   & 0.00\%   & 557.92 & 147.37\% & 238.22\% & 1    & D \\
 & $Obj_3$ & 0.3712 & 98.99\%  & 192.44 & 137.09\% & 97.38  & 0.00\%   & 169.10\% & 4200 & D \\
\midrule
 \multirow{3}{2cm}{Weighted sum} 
 & $Obj_1$ & 0.0956 & 0.68\%  & 167.43 & 119.28\% & 179.57 & 26.30\% & 122.14\% & 4209 & D  \\
 & $Obj_2$ & 0.3362 & 86.49\% & 0.00   & 0.00\%   & 406.72 & 98.99\% & 131.45\% & 3    & ND \\
 & $Obj_3$ & 0.2216 & 45.61\% & 111.99 & 79.78\%  & 101.84 & 1.43\%  & 91.91\%  & 4201 & ND \\
  \midrule
  \multirow{6}{2cm}{Lexicographic} 
 & $Obj_1, Obj_2$ & \multicolumn{9}{c}{\textit{No feasible solution found}} \\*	
 & $Obj_1, Obj_3$ & \multicolumn{9}{c}{\textit{No feasible solution found}} \\*	
 & $Obj_2, Obj_1$ & 0.3333 & 85.47\%  & 0.00 & 0.00\% & 409.89 & 100.00\% & 131.55\% & 1 & ND \\
 & $Obj_2, Obj_3$ & 0.3741 & 100.00\% & 0.00 & 0.00\% & 373.12 & 88.23\%  & 133.36\% & 2 & ND \\
 & $Obj_3, Obj_1$ & \multicolumn{9}{c}{\textit{No feasible solution found}}  \\*
 & $Obj_3, Obj_2$ & \multicolumn{9}{c}{\textit{No feasible solution found}}  \\	\midrule	
\multirow{6}{2cm}{Lexicographic with warm starts} 
 & $Obj_1, Obj_2$ & 0.0938 & 0.00\%   & 140.78 & 100.29\% & 225.96 & 41.14\%  & 108.40\% & 8400 & ND \\
 & $Obj_1, Obj_3$ & 0.0938 & 0.00\%   & 146.51 & 104.37\% & 225.96 & 41.14\%  & 112.19\% & 8402 & D  \\
 & $Obj_2, Obj_1$ & 0.3333 & 85.47\%  & 0.00   & 0.00\%   & 409.89 & 100.00\% & 131.55\% & 3    & ND \\
 & $Obj_2, Obj_3$ & 0.3741 & 100.00\% & 0.00   & 0.00\%   & 373.12 & 88.23\%  & 133.36\% & 1    & ND \\
 & $Obj_3, Obj_1$ & 0.2159 & 43.58\%  & 175.36 & 124.93\% & 97.38  & 0.00\%   & 132.31\% & 8406 & D  \\
 & $Obj_3, Obj_2$ & 0.2159 & 43.58\%  & 164.68 & 117.32\% & 97.38  & 0.00\%   & 125.15\% & 8407 & ND \\
\bottomrule
\end{tabular}
}
\end{sidewaystable}

\begin{sidewaystable}
\centering
\setlength{\tabcolsep}{4pt}
\renewcommand{\arraystretch}{0.7}
{\scriptsize
\caption{\green{Objectives range approximation for scenarios of BBCA\_2 and BBCA\_3 with classified waste.}\label{tab:obj_range_BBCAII_and_BBCAIII_clas}}
\begin{tabular}{c c| rrrrrrrrc}
\toprule
Method & \begin{tabular}{c} Optimized\\objective \end{tabular} & $Obj_1$ & $\Delta Obj_1$ (\%) & $Obj_2$ & $\Delta Obj_1$ (\%) & $Obj_3$ & $\Delta Obj_1$ (\%) & $L_2$ (\%)  & \begin{tabular}{r}Execution\\time (s)\end{tabular}  & Dominance \\ \midrule
\multicolumn{11}{c}{BBCA\_2} \\ \midrule
\multirow{3}{2.1cm}{Single objective optimization} 
 & $Obj_1$ & 0.0322 & 0.24\%   & 193.31 & 113.01\% & 205.11 & 43.45\%  & 121.08\% & 4201 & D \\
 & $Obj_2$ & 0.6056 & 167.57\% & 0.00   & 0.00\%   & 621.32 & 155.31\% & 228.48\% & 1    & D \\
 & $Obj_3$ & 0.1361 & 30.58\%  & 190.77 & 111.53\% & 43.45  & 0.00\%   & 115.64\% & 4200 & D \\ \midrule
 \multirow{3}{2cm}{Weighted sum} 
 & $Obj_1$ & 0.0314 & 0.00\%  & 159.80 & 93.42\% & 96.84  & 14.35\%  & 94.52\%  & 4274 & ND \\
 & $Obj_2$ & 0.3234 & 85.23\% & 0.00   & 0.00\%  & 415.52 & 100.00\% & 131.39\% & 4    & ND \\
 & $Obj_3$ & 0.0965 & 19.02\% & 135.33 & 79.12\% & 50.88  & 2.00\%   & 81.40\%  & 4200 & ND \\
  \midrule
  \multirow{6}{2cm}{Lexicographic} 
 & $Obj_1, Obj_2$ & \multicolumn{9}{c}{\textit{No feasible solution found}} \\*	
 & $Obj_1, Obj_3$ & \multicolumn{9}{c}{\textit{No feasible solution found}} \\*	
 & $Obj_2, Obj_1$ & 0.3234 & 85.23\% & 0.00 & 0.00\% & 415.52 & 100.00\% & 131.39\% & 1 & ND \\
 & $Obj_2, Obj_3$ & 0.3234 & 85.23\% & 0.00 & 0.00\% & 415.52 & 100.00\% & 131.39\% & 1 & ND \\*
 & $Obj_3, Obj_1$ & \multicolumn{9}{c}{\textit{No feasible solution found}}  \\*
 & $Obj_3, Obj_2$ & \multicolumn{9}{c}{\textit{No feasible solution found}}  \\	\midrule	
\multirow{6}{2cm}{Lexicographic with warm starts} 
 & $Obj_1, Obj_2$ & 0.0322 & 0.24\%  & 164.87 & 96.39\%  & 96.84  & 14.35\%  & 97.45\%  & 8410 & D  \\
 & $Obj_1, Obj_3$ & 0.0322 & 0.24\%  & 162.60 & 95.06\%  & 94.63  & 13.76\%  & 96.05\%  & 8482 & ND \\
 & $Obj_2, Obj_1$ & 0.3234 & 85.23\% & 0.00   & 0.00\%   & 415.52 & 100.00\% & 131.39\% & 3    & ND \\
 & $Obj_2, Obj_3$ & 0.3234 & 85.23\% & 0.00   & 0.00\%   & 415.52 & 100.00\% & 131.39\% & 2    & ND \\
 & $Obj_3, Obj_1$ & 0.0965 & 19.02\% & 180.43 & 105.48\% & 43.45  & 0.00\%   & 107.18\% & 8409 & D  \\
 & $Obj_3, Obj_2$ & 0.0965 & 19.02\% & 171.05 & 100.00\% & 43.45  & 0.00\%   & 101.79\% & 8401 & ND \\ \midrule
\multicolumn{11}{c}{BBCA\_3} \\ 
 \midrule
\multirow{3}{2.1cm}{Single objective optimization} 
 & $Obj_1$ & 0.0460 & 0.28\%   & 179.50 & 116.76\% & 511.22 & 118.50\% & 166.36\% & 4200 & D \\
 & $Obj_2$ & 0.6279 & 229.10\% & 0.00   & 0.00\%   & 659.36 & 158.09\% & 278.35\% & 1    & D \\
 & $Obj_3$ & 0.5948 & 216.10\% & 180.70 & 117.54\% & 67.84  & 0.00\%   & 246.08\% & 4265 & D \\
\midrule
 \multirow{3}{2cm}{Weighted sum} 
 & $Obj_1$ & 0.0453 & 0.00\%   & 143.69 & 93.47\% & 161.4  & 25.00\% & 96.76\%  & 4214 & ND \\
 & $Obj_2$ & 0.2996 & 100.00\% & 0.00   & 0.00\%  & 440.96 & 99.72\% & 141.22\% & 6    & ND \\
 & $Obj_3$ & 0.1264 & 31.92\%  & 149.94 & 97.53\% & 75.23  & 1.98\%  & 102.64\% & 4201 & ND \\
  \midrule
  \multirow{6}{2cm}{Lexicographic} 
 & $Obj_1, Obj_2$ & \multicolumn{9}{c}{\textit{No feasible solution found}} \\*	
 & $Obj_1, Obj_3$ & \multicolumn{9}{c}{\textit{No feasible solution found}} \\*	
 & $Obj_2, Obj_1$ & 0.2989 & 99.72\%  & 0.00 & 0.00\% & 442.01 & 100.00\% & 141.22\% & 1 & ND \\
 & $Obj_2, Obj_3$ & 0.2996 & 100.00\% & 0.00 & 0.00\% & 440.96 & 99.72\%  & 141.22\% & 3 & ND \\
 & $Obj_3, Obj_1$ & \multicolumn{9}{c}{\textit{No feasible solution found}}  \\*
 & $Obj_3, Obj_2$ & \multicolumn{9}{c}{\textit{No feasible solution found}}  \\	\midrule	
\multirow{6}{2cm}{Lexicographic with warm starts} 
 & $Obj_1, Obj_2$ & 0.0460 & 0.28\%   & 139.96 & 91.04\%  & 172.16 & 27.88\%  & 95.22\%  & 8405 & ND \\
 & $Obj_1, Obj_3$ & 0.0460 & 0.28\%   & 140.00 & 91.07\%  & 172.16 & 27.88\%  & 95.24\%  & 8490 & D  \\
 & $Obj_2, Obj_1$ & 0.2989 & 99.72\%  & 0.00   & 0.00\%   & 442.01 & 100.00\% & 141.22\% & 4    & ND \\
 & $Obj_2, Obj_3$ & 0.2996 & 100.00\% & 0.00   & 0.00\%   & 440.96 & 99.72\%  & 141.22\% & 4    & ND \\
 & $Obj_3, Obj_1$ & 0.1336 & 34.75\%  & 164.10 & 106.75\% & 67.84  & 0.00\%   & 112.26\% & 8401 & D  \\
 & $Obj_3, Obj_2$ & 0.1329 & 34.46\%  & 153.73 & 100.00\% & 67.84  & 0.00\%   & 105.77\% & 8402 & ND \\
\bottomrule
\end{tabular}
}
\end{sidewaystable}

\end{document}